\documentclass[amssymb,amsmath,prd,twocolumn,superscriptaddress,nofootinbib]{revtex4-1}

\usepackage{graphicx}
\usepackage{bm}
\usepackage{amssymb,amsmath}
\usepackage{mathrsfs}
\usepackage{latexsym}
\usepackage{color}
\usepackage[normalem]{ulem} 
\usepackage{dcolumn}
\usepackage[colorlinks=true,citecolor=blue,urlcolor=blue]{hyperref}
\usepackage[usenames,dvipsnames]{xcolor}

\allowdisplaybreaks

\newcommand{\CITA}{\affiliation{Canadian Institute for Theoretical Astrophysics, 60 St. George Street, Toronto, Ontario, M5S 3H8, Canada}}

\definecolor{azgreen}{rgb}{0.03,0.47,0.19}
\definecolor{kcmagenta}{rgb}{0.54, 0.17, 0.88}
\definecolor{chorange}{rgb}{0.851, 0.372, 0.007}

\begin{document}

\title{Measuring the neutron star tidal deformability with equation-of-state-independent relations and gravitational waves}

\author{Katerina Chatziioannou}\CITA
\author{Carl-Johan Haster}\CITA
\author{Aaron Zimmerman}\CITA

\date{\today}

\begin{abstract}
Gravitational wave measurements of binary neutron star coalescences offer information about the properties of the extreme matter that comprises the stars. 
Despite our expectation that all neutron stars in the Universe obey the same equation of state, i.e.~the properties of the matter that forms them are universal, current tidal inference analyses treat the two bodies as independent. 
We present a method to measure the effect of tidal interactions in the gravitational wave signal -- and hence constrain the equation of state -- that assumes that the two binary components obey the same equation of state. 
Our method makes use of a relation between the tidal deformabilities of the two stars given the ratio of their masses, a relation that has been shown to only have a weak dependence on the equation of state. 
We use this to link the tidal deformabilities of the two stars in a realistic parameter inference study while simultaneously marginalizing over the error in the relation. 
This approach incorporates more physical information into our analysis, thus leading to a better measurement of tidal effects in gravitational wave signals. 
Through simulated signals we estimate that uncertainties in the measured tidal parameters are reduced by a factor of at least 2 -- and in some cases up to 10 -- depending on the equation of state and mass ratio of the system.
\end{abstract}

\maketitle
  
\section{Introduction}

The recent detection of a binary neutron star (BNS) coalescence with gravitational waves (GWs)~\cite{TheLIGOScientific:2017qsa} by the Advanced LIGO~\cite{TheLIGOScientific:2014jea} and Advanced Virgo~\cite{TheVirgo:2014hva} detectors offers a new way to study the properties of dense matter~\cite{Lattimer:2015nhk,Ozel:2016oaf}. 
The equation of state (EoS) of the supranuclear, cold matter that forms NSs leaves its imprint on the GW signal during the late stages of the inspiral~\cite{Blanchet:2014zz} when the two bodies are tidally distorted, and during the merger and postmerger phases, e.g.~\cite{Xing:1994ak,Shibata:2005xz,Flanagan:2007ix,Read:2009yp} and Refs.~\cite{Faber2012,Baiotti:2016qnr,Paschalidis:2016vmz} for reviews.

The initial tidal distortion, before the eventual disruption of coalescing NSs by strong tidal fields caused by their companion, can be quantified through the \emph{tidal deformability}~\cite{Flanagan:2007ix,Hinderer:2007mb,Hinderer:2009ca,Vines:2011ud}. 
To leading multipolar order this parameter is defined through $Q_{ij} = -\lambda E_{ij}$, where $Q_{ij}$ is the induced mass quadrupole moment on a star that is subject to an electric tidal field $E_{ij}$, and $\lambda$ is the tidal deformability.\footnote{Throughout this work we use geometric units where $G=c=1$.}
With these, ten intrinsic physical parameters characterize the binary: the masses $m_i$, the spin angular momenta $\mathbf{S_i}$, and the dimensionless tidal deformabilities $\Lambda_i\equiv \lambda_im_i^5$ of the two NSs, where $i \in \{1,2\}$ and $i=1$ always corresponds to the more massive object. 
A number of studies have exploited the fact that the tidal deformability depends on the NS EoS to conclude that it is possible to put constraints on the NS radius and EoS through GW measurements~\cite{Read:2009yp,Pannarale:2011pk,PhysRevD.85.123007,PhysRevD.85.044061,Read:2013zra,Maselli:2013rza,Lackey:2013axa,DelPozzo:2013ala,Agathos:2015uaa,Wade:2014vqa,Lackey:2014fwa,Hotokezaka:2016bzh,Chatziioannou:2015uea} or by combining electromagnetic and GW information~\cite{Radice:2017lry,Margalit:2017dij,Bauswein:2017vtn,2018ApJ...852L..25R}.

However, unlike the masses and the spins of the two binary components, their tidal deformabilities are not independent. 
For a given EoS, the dimensionless tidal deformability is a function of the mass of the NS, and this function has certain properties. 
For example, all viable, realistic EoS models proposed to date predict dimensionless tidal deformabilties that are a decreasing function of the NS mass for stable NS configurations. 
In the context of a BNS this means that for a given EoS the most massive component will have the smallest $\Lambda_i$. 

This additional \emph{physical modeling} of the system, however, has traditionally not been straightforward to incorporate in a realistic parameter inference analysis, as it would require an efficient parametrization of the $\Lambda(m)$ relation that is applicable to a large range of possible EoSs. 
Such a parametrization was recently put forward by Yagi and Yunes~\cite{Yagi:2015pkc}, who proposed a relation between the symmetric combination $\Lambda_s \equiv (\Lambda_2+\Lambda_1)/2$, the antisymmetric combination $\Lambda_a \equiv (\Lambda_2-\Lambda_1)/2$, and the mass ratio of the system $q \equiv m_2/m_1\leq1$ that depends only weakly on the EoS. 
The resulting relation $\Lambda_a(\Lambda_s,q;\vec{b})$ was shown to reproduce realistic EoSs with a relative error of $\lesssim 10\%$. 
The quantities $\vec{b}$ are parameters whose values and errors are determined by fitting to a range of EoSs~\cite{Yagi:2015pkc}.

\begin{figure}[t]
\includegraphics[width=\columnwidth,clip=true]{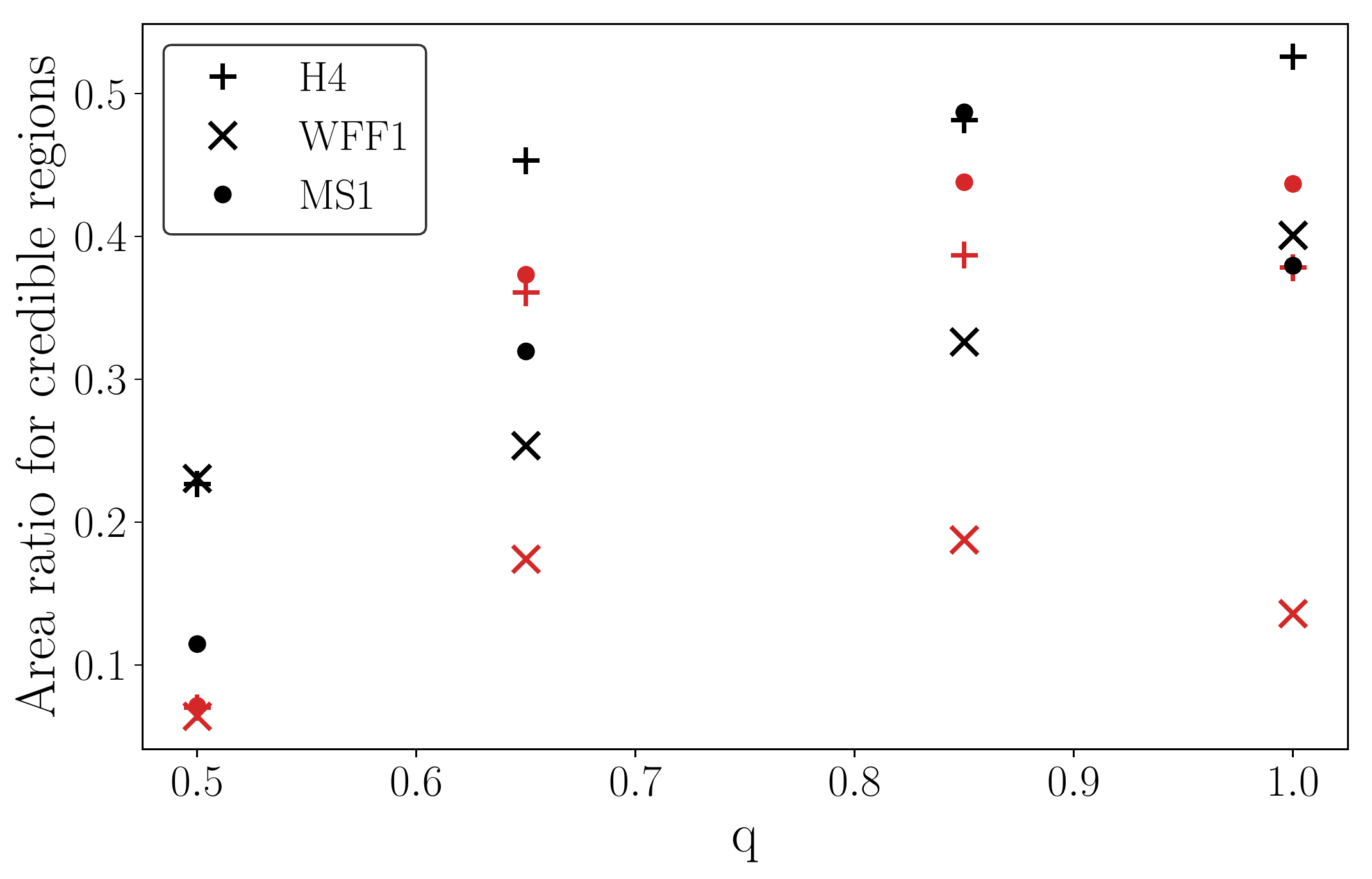}\
\caption{ \label{fig:ratio} Ratio of the areas of the 90\% credible intervals of the $\Lambda_1-\Lambda_2$ two-dimensional posterior probability density when the EoS-independent relation is imposed in the analysis and when the individual tidal deformabilities are presumed to be independent for different EoSs, mass ratios, and an SNR of 30 (black) and 15 (red). The method proposed here leads to a reduction in the measurement error for $\Lambda_1-\Lambda_2$ by a factor between 2 and 10. }
\end{figure}

In this study we use the EoS-independent parametrization of Ref.~\cite{Yagi:2015pkc} to study how well we can measure the tidal deformability of NSs and put constraints on their EoS. 
We simulate GW signals as observed by the LIGO and Virgo detectors and employ realistic parameter estimation to infer the parameters of the system by imposing the $\Lambda_a(\Lambda_s,q;\vec{b})$ relation, while marginalizing over its intrinsic error. 
In Fig.~\ref{fig:ratio} we show the ratio of the areas of the $90\%$ credible regions for the measurement  of $\Lambda_1-\Lambda_2$ when we employ the EoS-independent relation and when we assume that the individual tidal deformabilities are independent. 
We present results for three representative EoSs (H4, WFF1, MS1), two values for the signal-to-noise ratio (SNR) of the system,  and as a function of the mass ratio of the system. 
In all cases use of the EoS-independent relation to link $\Lambda_1$ and $\Lambda_2$ leads to a reduction in the credible region by factors of 2 to 10.

\begin{figure}[t]
\includegraphics[width=\columnwidth,clip=true]{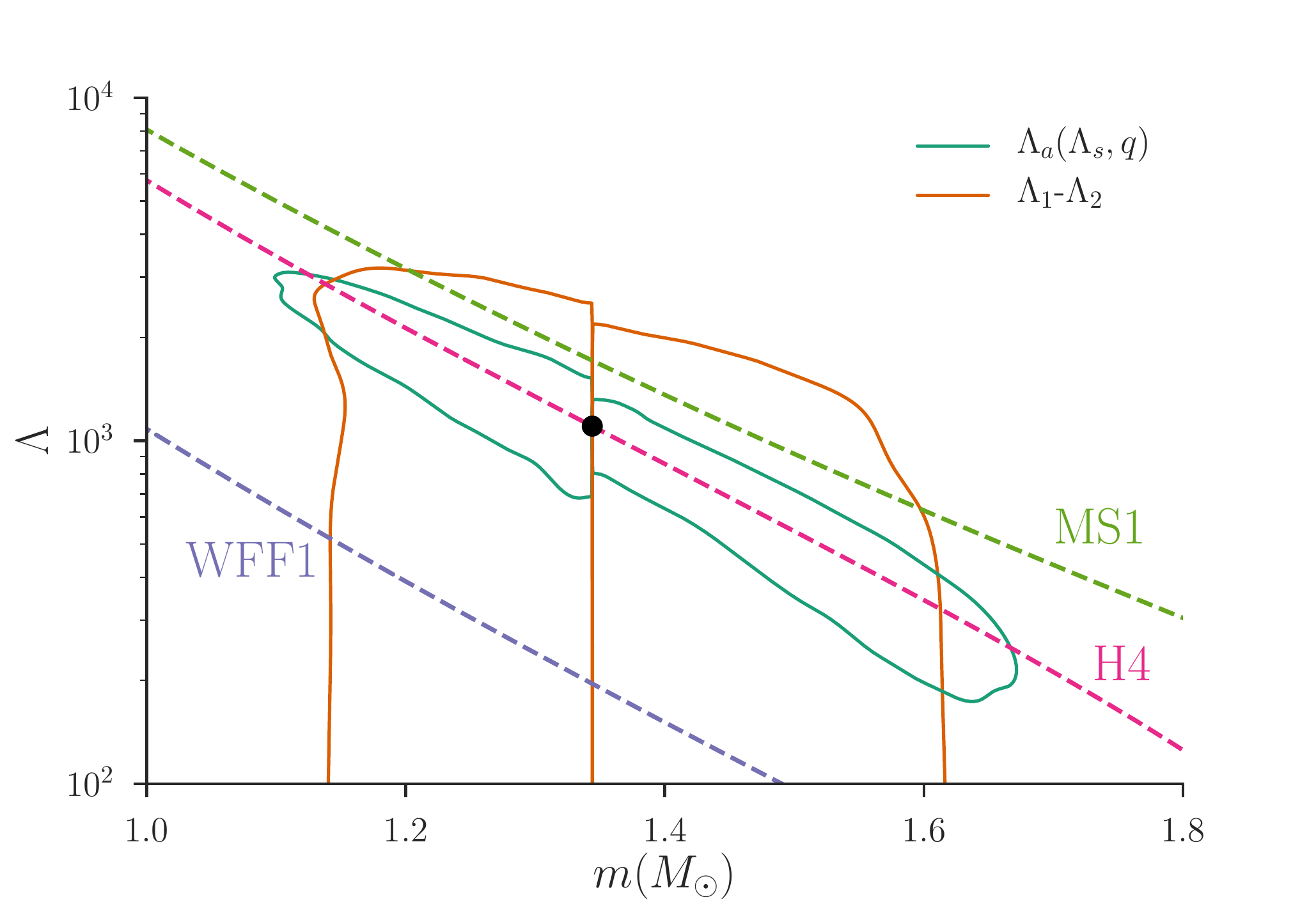}
\caption{ \label{fig:EoSs} The 90\% contours of the marginalized posterior probability density for $\Lambda_i$ and $m_i$ for each NS when using the EoS-independent relation (green contours) and when the components' tidal deformabilties are independent (orange contours) for a signal created with the H4 EoS, an SNR of 30 and a mass ratio of 1. 
Overplotted is the dimensionless tidal deformability as a function of the mass for the 3 EoS models studied here. 
All EoSs are consistent with the posteriors derived under independent tidal deformabilties at the 90\% level, but only H4 (the correct EoS) is consistent with the data if we use the EoS-independent relation.}
\end{figure}

As a result of the increased measurement accuracy, the relation presented here can result in an improved EoS determination from GW data. 
Figure~\ref{fig:EoSs} shows the two-dimensional posterior density for $m_i - \Lambda_i$ for a signal created with the H4 EoS. 
In orange we show 90\% contours from an analysis that assumes that the individual tidal deformabilities are independent, while in green we show the result of imposing the EoS-independent relation, while marginalizing over its error. 
Contours on the left side of the plot correspond to the lighter binary component, while contours on the right side refer to the heavier star. 
For reference we overplot the prediction from the three EoSs studied here and show that only the true EoS is consistent with the data at the 90\% level once the EoS-independent relation is invoked. 
Given its physical applicability and improved efficiency in the measurement of tides, we propose that this method be used for the analysis of BNS signals.

The rest of the paper provides the details of this study. 
In Sec.~\ref{analysis}, we describe the EoS-independent relation we employ, as well as the simulated GW signals we use to study it. 
In Sec.~\ref{results}, we present and discuss the results of performing parameter estimation on these signals. 
In Sec.~\ref{conclusions} we conclude.

\section{Analysis}
\label{analysis}

All stable NSs are expected to obey the same EoS describing the properties of their dense, cold matter. 
In this section we describe a proposed relation between the tidal deformabilities of two stars in a binary that depends only weakly on the EoS. 
We also provide details about the simulated BNS signals which we use to study the relation, as well as the parameter estimation techniques we employ.

\subsection{EoS-independent relations}

Many NS properties, such as the radius, the moment of inertia, or the maximum mass are determined by the EoS and are thus uncertain. 
However, it has been discovered that certain properties are related to each other in ways that are not strongly dependent on the specific EoS, giving rise to EoS-independent relations (see Ref.~\cite{Yagi:2016bkt} for a review and references).
Among them, of particular interest in the context of GW signals from BNSs is a relation between the dimensionless tidal deformabilities of the two NSs and their mass ratio. 
The relation can be expressed as~\cite{Yagi:2015pkc}
\begin{align}
\Lambda_a(\Lambda_s,q;\vec{b})&=F_n(q)\Lambda_s\frac{1+\sum_{i=1}^{3}\sum_{j=1}^{2}b_{ij} q^j\Lambda_s^{-i/5}}{1+\sum_{i=1}^{3}\sum_{j=1}^{2}c_{ij} q^j\Lambda_s^{-i/5}},\label{BL}
\\
F_n(q)&=\frac{1-q^{10/(3-n)}}{1+q^{10/(3-n)}},
\end{align}
where the parameters $\vec{b}=\{b_{ij},c_{ij}\}$ are determined through fitting to realistic EoSs and $n=0.743$ is the average polytropic index for these EoSs. The values of $\vec{b}=\{b_{ij},c_{ij}\}$ are given in Table~\ref{table:fittingparameters}.

\begin{table}[t]
\begin{centering}
\begin{tabular}{c|cc||cc|cc}
\noalign{\smallskip}
\hline
\noalign{\smallskip}
$b_{11}$  && -27.7408  && $c_{11}$   && -25.5593 \\
$b_{12}$  && 8.42358 && $c_{12}$  && 5.58527\\
$b_{21}$ 	&& 122.686   && $c_{21}$     && 92.0337 \\
$b_{22}$  &&  -19.7551   && $c_{22}$     && 26.8586\\
$b_{31}$  && -175.496   && $c_{31}$     &&  -70.247\\
$b_{32}$  &&  133.708  && $c_{32} $    &&  -56.3076\\
\noalign{\smallskip}
\hline
\hline
\end{tabular}
\end{centering}
\caption{Fitting parameters $\vec{b}$ for Eq.~\eqref{BL}. The numerical values for the parameters are different than the ones given in Ref.~\cite{Yagi:2015pkc} since we have factored out their parameter $a$. 
We have verified that the resulting EoS-independent relation is the same.}
\label{table:fittingparameters}
\end{table}
\begin{figure}[t]
\includegraphics[width=\columnwidth,clip=true]{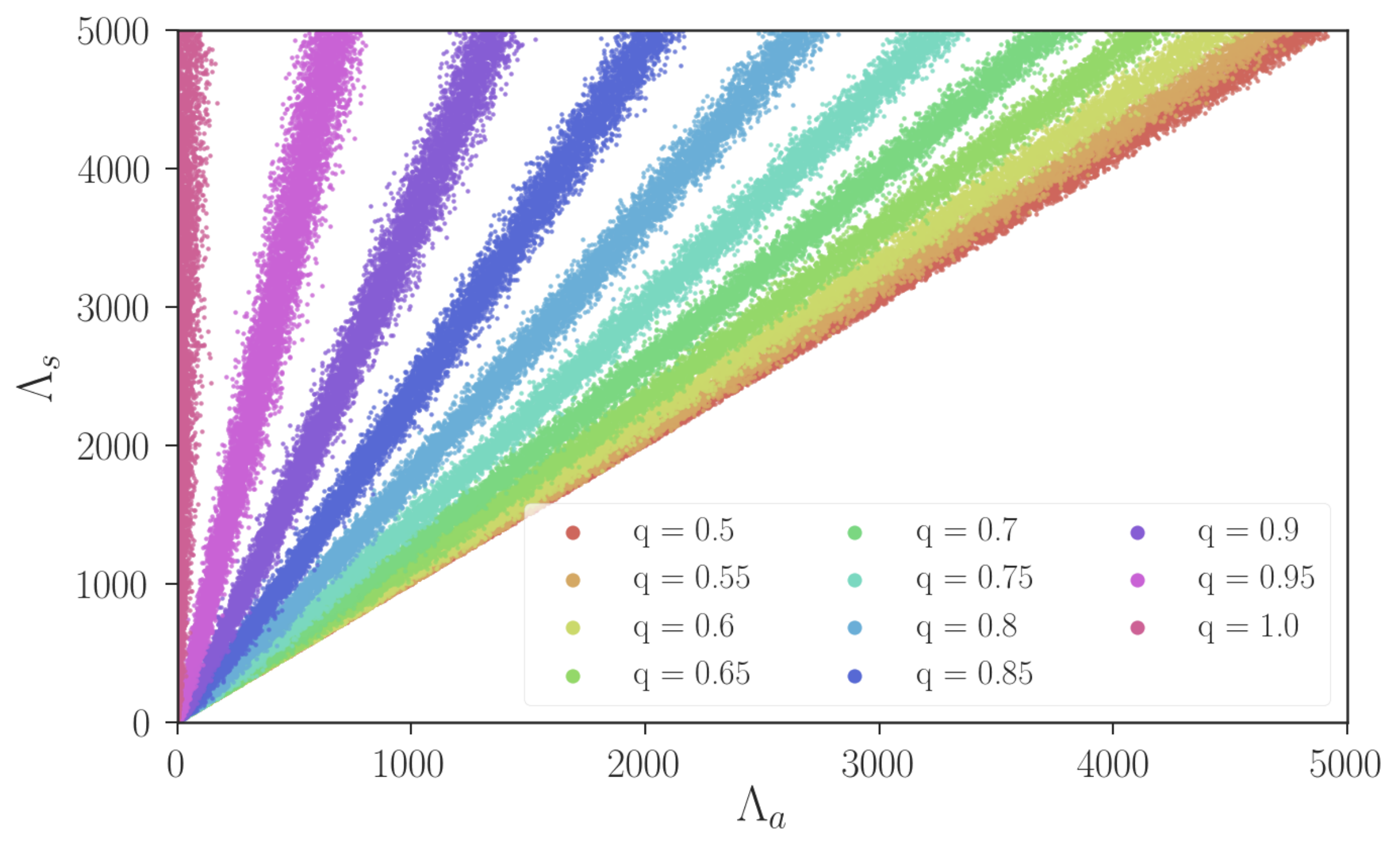}\\
\includegraphics[width=\columnwidth,clip=true]{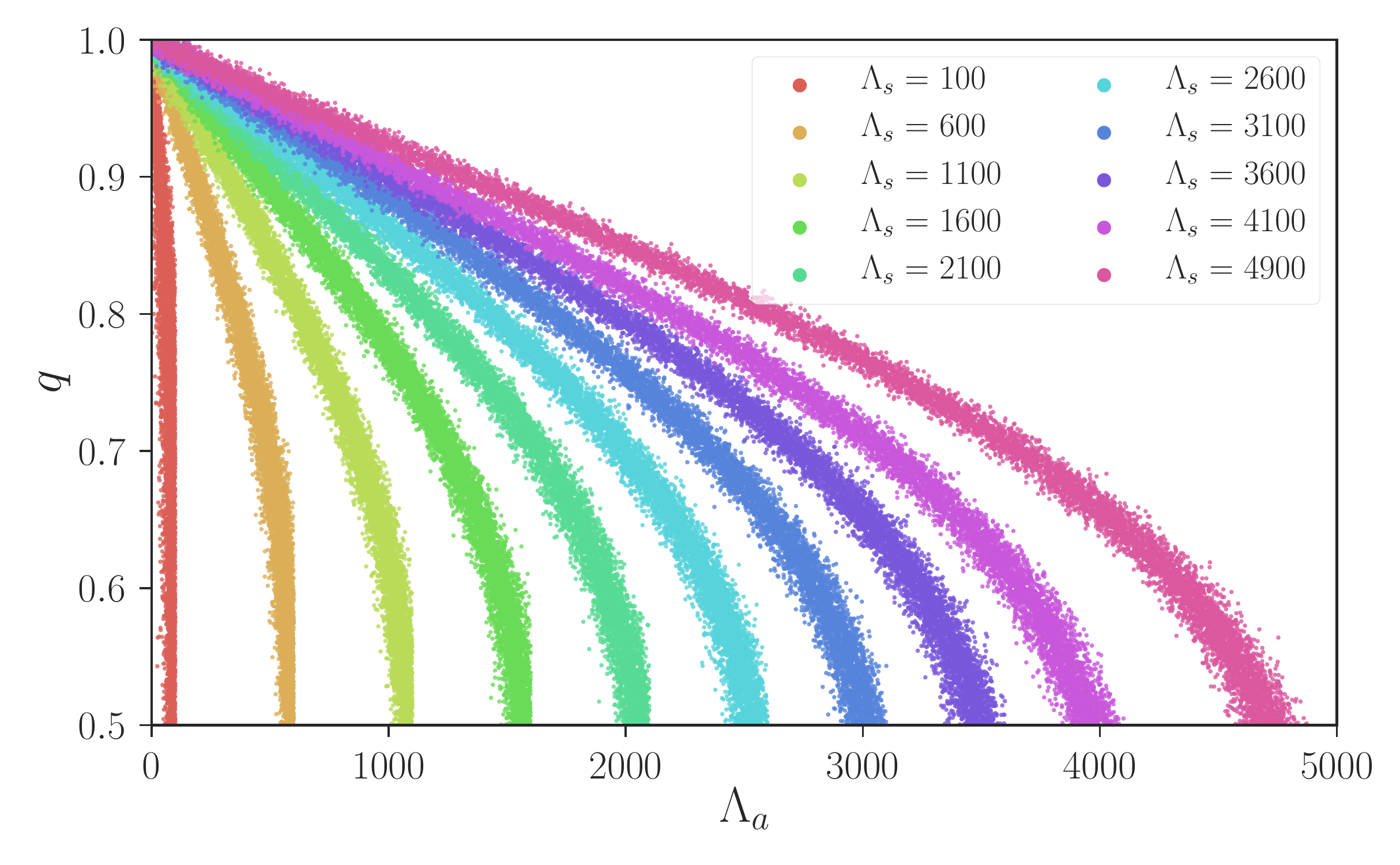}
\caption{ \label{fig:relation} EoS-independent relation between the antisymmetric and the symmetric combination of the tidal deformabilities (top panel) for different mass ratios, and the mass ratio for different values of the symmetric tidal deformability (bottom panel). 
The scatter around the EoS-independent relation represents its $\lesssim10\%$ relative error.}
\end{figure}

This relation has been shown to reproduce the dimensionless tidal deformabilties predicted by realistic EoSs to a relative error of $\lesssim10\%$. 
Figure~\ref{fig:relation} offers a graphical representation of the relation. 
The top panel shows $\Lambda_a(\Lambda_s)$ for various values of the mass ratio while the bottom panel shows $\Lambda_a(q)$ for various values of $\Lambda_s$. 
The scatter is a depiction of the error in the EoS-independent relation. When the binary components have equal masses $(q=1)$ the two NSs also have the same tidal deformabilities, leading to $\Lambda_1=\Lambda_2$ and $\Lambda_a=0$.
Despite the accuracy of the relation, its residual error can cause biased parameter inference if not properly taken into account~\cite{Chatziioannou:2017ixj}. 
In the context of parameter estimation for BNS GW signals, we use this relation to link the dimensionless tidal deformabilities of the two bodies, while we marginalize over the residual error in the relation with a procedure that we describe in Sec.~\ref{sec:marg}.

Use of this EoS-independent relation not only enables us to include physically motivated constraints in our inference, but also allows for better measurement of the individual tidal deformabilities. 
It is well known that, similar to the chirp mass ${\cal{M}}\equiv (m_1 m_2)^{3/5}/(m_1+m_2)^{1/5}$, there is a single tidal parameter measured most accurately with GWs~\cite{Wade:2014vqa}
\begin{equation}
\tilde{\Lambda}\equiv \frac{16}{13} \frac{(m_1+12m_2)m_1^4\Lambda_1+(m_2+12m_1)m_2^4\Lambda_2}{(m_1+m_2)^5}.
\end{equation}
This parameter enters the waveform phase at 5PN~\cite{Flanagan:2007ix} order\footnote{An N-PN order term is proportional to $(u/c)^{2N}$ relative to the leading order term, where $c$ is the speed of light and $u$ some characteristic velocity of the system.} and has the largest EoS-related effect on the signal. 
An independent tidal parameter enters the phase at 6PN order~\cite{Vines:2011ud} but its effect is subdominant~\cite{Wade:2014vqa}. Similarly the spin-induced quadrupole moment also depends on the EoS and enters at 2PN~\cite{Poisson:1997ha} but it has a small effect on the waveform despite its low PN order due to its small magnitude and correlation with the spins.

Despite being the best measured tidal parameter, $\tilde{\Lambda}$ has limited astrophysical interest and it is the $\Lambda_i$ of the two binary components that have the potential to reveal the NS EoS. However, it is hard to place constraints on the individual tidal deformabilities of the two stars if we have only measured one tidal parameter in the form of $\tilde{\Lambda}$. 
The EoS-independent relation employed here offers a way to extract the tidal deformabilities of both binary components since they are linked to each other through the mass ratio of the system.

\subsection{Error marginalization}
\label{sec:marg}

Despite its high degree of accuracy, any residual error in the EoS-independent relation $\Lambda_a=\Lambda_a(\Lambda_s,q;\vec{b})$ can jeopardize inference about the correct EoS. 
In our analysis, we marginalize over intrinsic error in the relation by studying its residuals. 
We compute $\Lambda_a^{\rm{true}}$ predicted by various realistic EoSs for different mass ratios and compare them to the prediction by the EoS-independent relation $\Lambda_a^{\rm{relation}}$ for the same mass ratios and values of $\Lambda_s$. 
We conservatively assume that the residuals $\Lambda_a^{\rm{relation}}-\Lambda_a^{\rm{true}}$ obey a Gaussian distribution with a mean $\mu(\Lambda_s,q)$ and standard deviation $\sigma(\Lambda_s,q)$ and fit for the mean and standard deviation to obtain
\begin{align}
\mu(\Lambda_s,q)&= \frac{\mu_{\Lambda_s}(\Lambda_s)+\mu_q(q)}{2},
\\
\sigma^2(\Lambda_s,q)& = \sigma^2_{\Lambda_s}(\Lambda_s)+\sigma^2_q(q),
\end{align}
where
\begin{align}
\mu_{\Lambda_s}(x)&=\frac{\mu_1}{x^2}+\frac{\mu_2}{x}+\mu_3,\label{mu-begin} 
\\
\mu_q(x)&= \mu_4x^2+\mu_5x+\mu_6,
\\
\sigma_{\Lambda_s}(x)&= \sigma_1x\sqrt{x}+\sigma_2x+\sigma_3\sqrt{x}+\sigma_4,
\\
\sigma_q(x)&= \sigma_5 x^2+\sigma_6x+ \sigma_7.\label{mu-end}
\end{align}
The fitting coefficients $\mu_i$ and $\sigma_i$ are provided in Table~\ref{table:fittingparameters2}. 

\begin{table}[t]
\begin{centering}
\begin{tabular}{c|cc||cc|cc}
\noalign{\smallskip}
\hline
\noalign{\smallskip}
$\mu_1$  && 137.1252739  && $\sigma_1$   && -0.0000739 \\
$\mu_2$  && -32.8026613 && $\sigma_2$  &&0.0103778\\
$\mu_3$ 	&& 0.5168637   && $\sigma_3$     && 0.4581717 \\
$\mu_4$  &&  -11.2765281   && $\sigma_4$     && - 0.8341913\\
$\mu_5$  && 14.9499544   && $\sigma_5$     &&  -201.4323962\\
$\mu_6$  &&  -4.6638851  && $\sigma_6$    && 273.9268276 \\
  &&    && $\sigma_7$    &&-71.2342246 \\
\noalign{\smallskip}
\hline
\hline
\end{tabular}
\end{centering}
\caption{Fitting coefficients of Eqs.~\eqref{mu-begin}--\eqref{mu-end} for the relative error in the EoS-independent relation. }
\label{table:fittingparameters2}
\end{table}

In practice, we select a value of $q$ and $\Lambda_s$ in the parameter estimation sampling algorithm and then we compute $\Lambda_a$ though

\begin{equation}
\Lambda_a=\Lambda_a(\Lambda_s,q;\vec{b})+{\cal{N}}(\mu(\Lambda_s,q),\sigma(\Lambda_s,q)),
\end{equation}
where ${\cal{N}}(\mu,\sigma)$ is a normal distribution with mean $\mu$ and standard deviation $\sigma$. We note that this error marginalization procedure ensures the recovery of unbiased posterior even for extremely loud signals. In these cases, the statistical measurement uncertainty of tidal parameters is negligible and the overall uncertainty budget is dominated by the systematic uncertainty due to the EoS-independent relation. In other words, use of the EoS-independent relation limits the measurement uncertainty of the tidal parameters to be above $\sim10\%$ regardless of the signal strength.

In Sec.~\ref{results} we use the procedure described above to marginalize over the intrinsic error in the EoS-independent relation and show that we retrieve unbiased measurements of the tidal deformabilities.

\subsection{Simulated signals} 
\label{signals}

To quantify the improvement in the measurement of the tidal deformability due to EoS-independent relations we study simulated GW signals emitted from BNS coalescences. 
We assume that the signals are detected by a network of two LIGOs and Virgo at design sensitivity~\cite{Aasi:2013wya} with no calibration uncertainties or added Gaussian noise. 
For simplicity we study nonspinning systems with a detector frame chirp mass of $1.17M_{\odot}$ and $4$ different values of the mass ratio $q\in \{1,0.85,0.65,0.5 \}$. 
We choose the systems' orientation such that the orbital angular momentum points towards the detectors and the systems are located directly overhead of the LIGO--Livingston detector. 
We scale the distance to the source such that we achieve an SNR of  either 15 or 30 and present our results in terms of the SNR.

We select the waveform model {\tt IMRPhenomD\_NRTidal} both for the generation and for the analysis of the signals. 
{\tt IMRPhenomD\_NRTidal} is based on the phenomenological inspiral-merger-ringdown model {\tt IMRPhenomD} originally constructed for BBH systems and assuming that the objects' spins remain aligned with the orbital angular momentum~\cite{Husa:2015iqa,Khan:2015jqa}.\footnote{
At the outset of this study, no waveform model included both the effects of spin-precession and tidal effects calibrated to numerical relativity results. Such a model has recently become available~\cite{Dietrich:2018uni}, and we expect that its use in future work will improve the measurement of the masses and the spins of the stars~\cite{Chatziioannou:2014bma,Chatziioannou:2014coa,Farr:2015lna}.} 
Tidal effects are then included on top of the BBH waveform~\cite{Barkett:2015wia}, calibrated through BNS numerical relativity simulations~\cite{Dietrich:2017aum}. 
The resulting model~\cite{Dietrich:2018uni} has been used for the analysis of GW170817~\cite{TheLIGOScientific:2017qsa}.

\begin{figure}[t]
\includegraphics[width=0.9\columnwidth,clip=true]{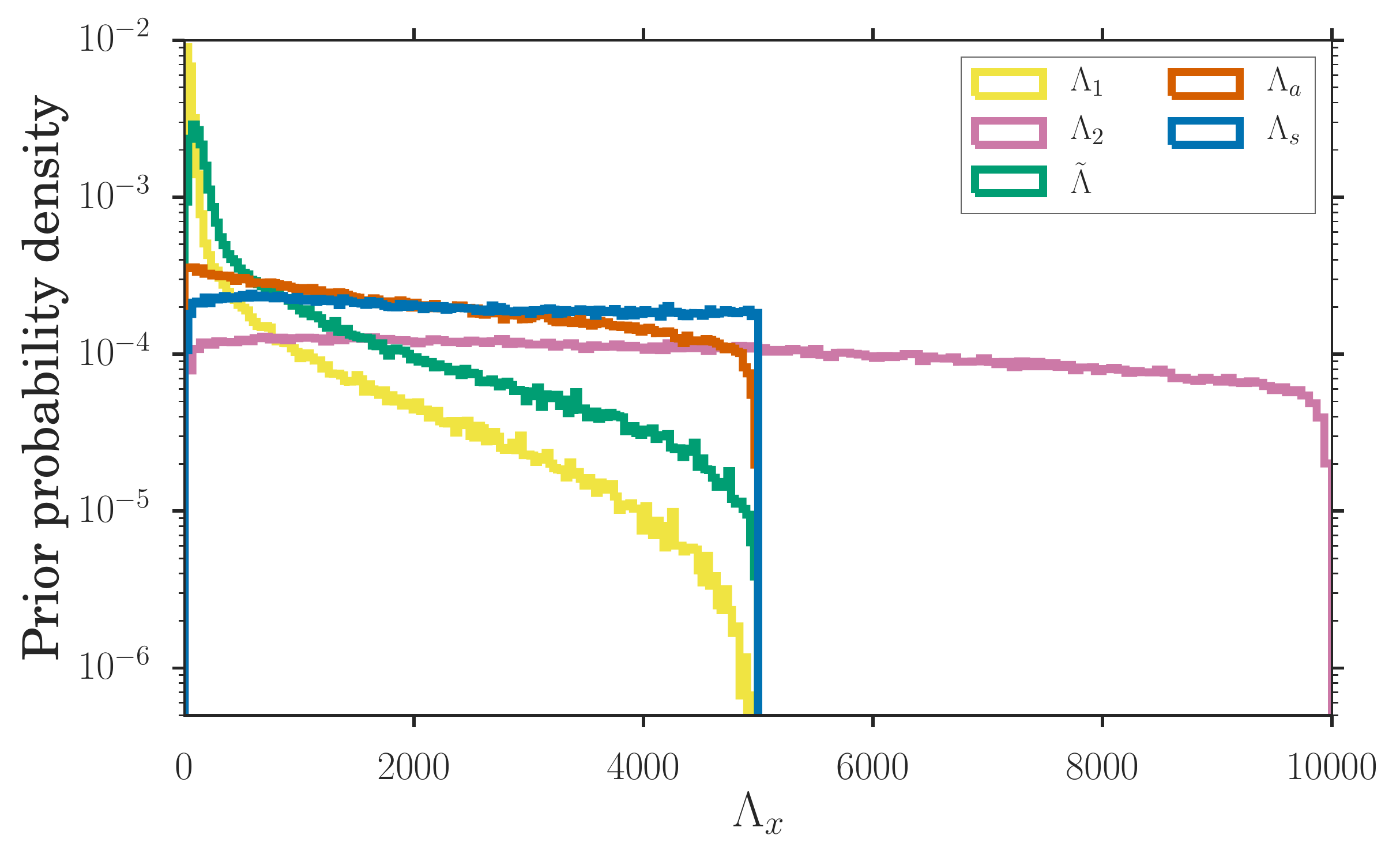}\\
\includegraphics[width=0.9\columnwidth,clip=true]{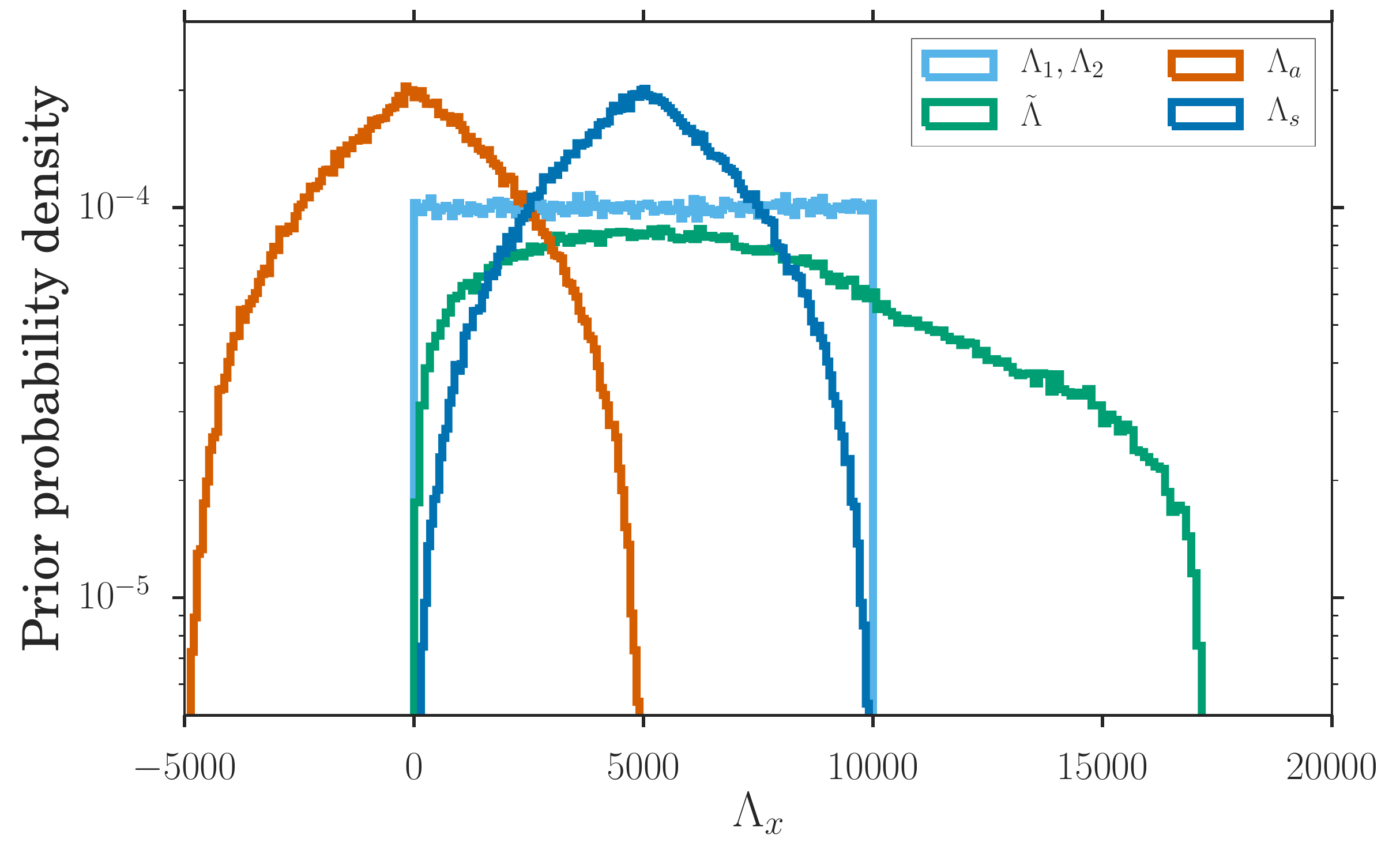}\\
\caption{ \label{fig:priors} Prior probability density distributions for the various dimensionless tidal deformability parameters. 
The top panel shows the prior densities when employing the EoS-independent relation. 
The $\Lambda_s$ prior is chosen to be uniform between $0$ and $5000$, while the prior of the other tidal parameters is informed by the $\Lambda_a=\Lambda_a(\Lambda_s,q;\vec{b})$ relation. 
The prior on the mass ratio is the result of flat priors on the individual masses.
The bottom panel shows the prior densities for the analyses without the EoS-independent relation included.
Here the $\Lambda_1$ and $\Lambda_2$ priors are chosen to be uniform between $0$ and $10000$, which in turn defines the remaining tidal parameters.}
\end{figure}

We then use the Bayesian Inference code {\tt LALInference}~\cite{Veitch:2014wba} to sample the joint posterior distribution of the system parameters. We assume flat priors on the component masses. 
For the dimensionless spin components along the orbital angular momentum, $\chi_{z,i} = S_{z,i}/m_i^2$, we use a prior distribution which results from assuming spins uniform in magnitude within the range $(0,0.05)$ and isotropic in direction; projecting these spins onto the orbital angular momentum gives a prior for $\chi_{z,i}$ peaked around zero.
The priors for the dimensionless tidal deformabilities differ according to the assumptions we make about them.
When we do not assume any relation between the components' dimensionless tidal deformabilities we choose flat, and independent, priors for each $\Lambda_i$ in $(0,10000)$. 
When, on the other hand, we employ the $\Lambda_a=\Lambda_a(\Lambda_s,q;\vec{b})$ EoS-independent relation, we use a prior uniform in $\Lambda_s$, now in $(0,5000)$ with the additional constraints of $0 \le \Lambda_1 \le \Lambda_2$. 
In Fig.~\ref{fig:priors}, we show the priors for the various tidal deformability parameters.

We study three EoSs with different predictions for the radii of NSs: 
(i) WFF1~\cite{PhysRevC.38.1010} is a soft EoS resulting in NS with a radius of about $10$ km; 
(ii) H4~\cite{Lackey:2005tk} is a stiffer EoS resulting in NS with a radius of about $14$ km. 
It is marginally consistent with GW170817 at the 90\% level~\cite{TheLIGOScientific:2017qsa}; 
(iii) MS1~\cite{Mueller:1996pm,Read:2008iy} is a stiff EoS resulting in NS with a radius of about $14.5$ km and it is inconsistent with GW170817. 
Even though MS1 it has been ruled out observationally and WFF1 has been suggested to be inconsistent with nuclear calculations~\cite{Hebeler:2013nza}, we use them as extreme examples of stiff and soft EoSs. 
The dimensionless tidal deformability as a function of the NS mass for these three EoSs is presented in Fig.~\ref{fig:EoSs}.

\section{Results}
\label{results}

Each simulated signal is analyzed with and without the EoS-independent relation between the tidal deformabilities of the two stars. 
In this section we present our results and demonstrate that the use of the EoS-independent relation improves the measurement of tides.

\subsection{Measurement of tidal deformabilities}

\begin{figure*}
\includegraphics[width=0.5\columnwidth,clip=true]{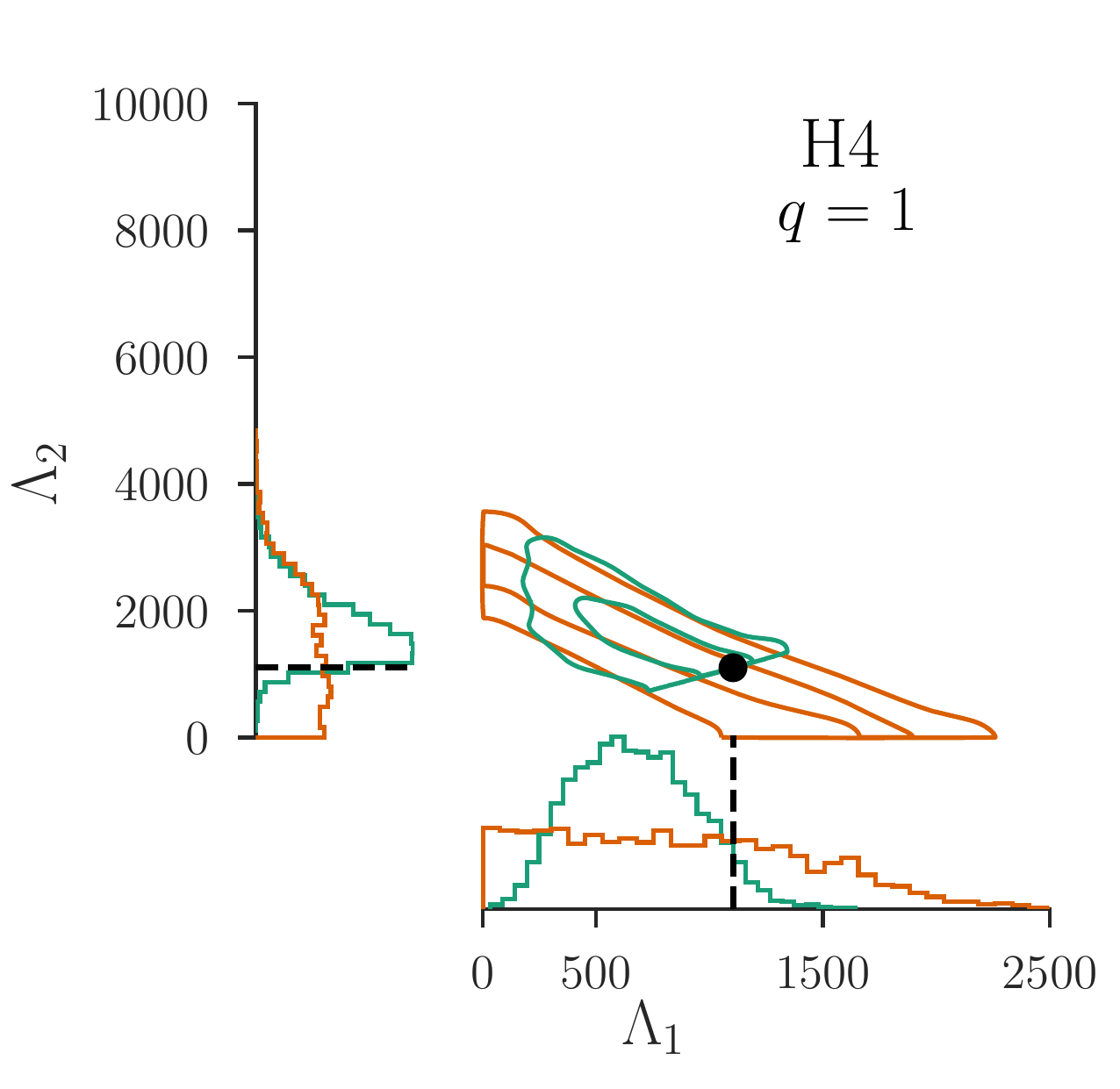}
\includegraphics[width=0.5\columnwidth,clip=true]{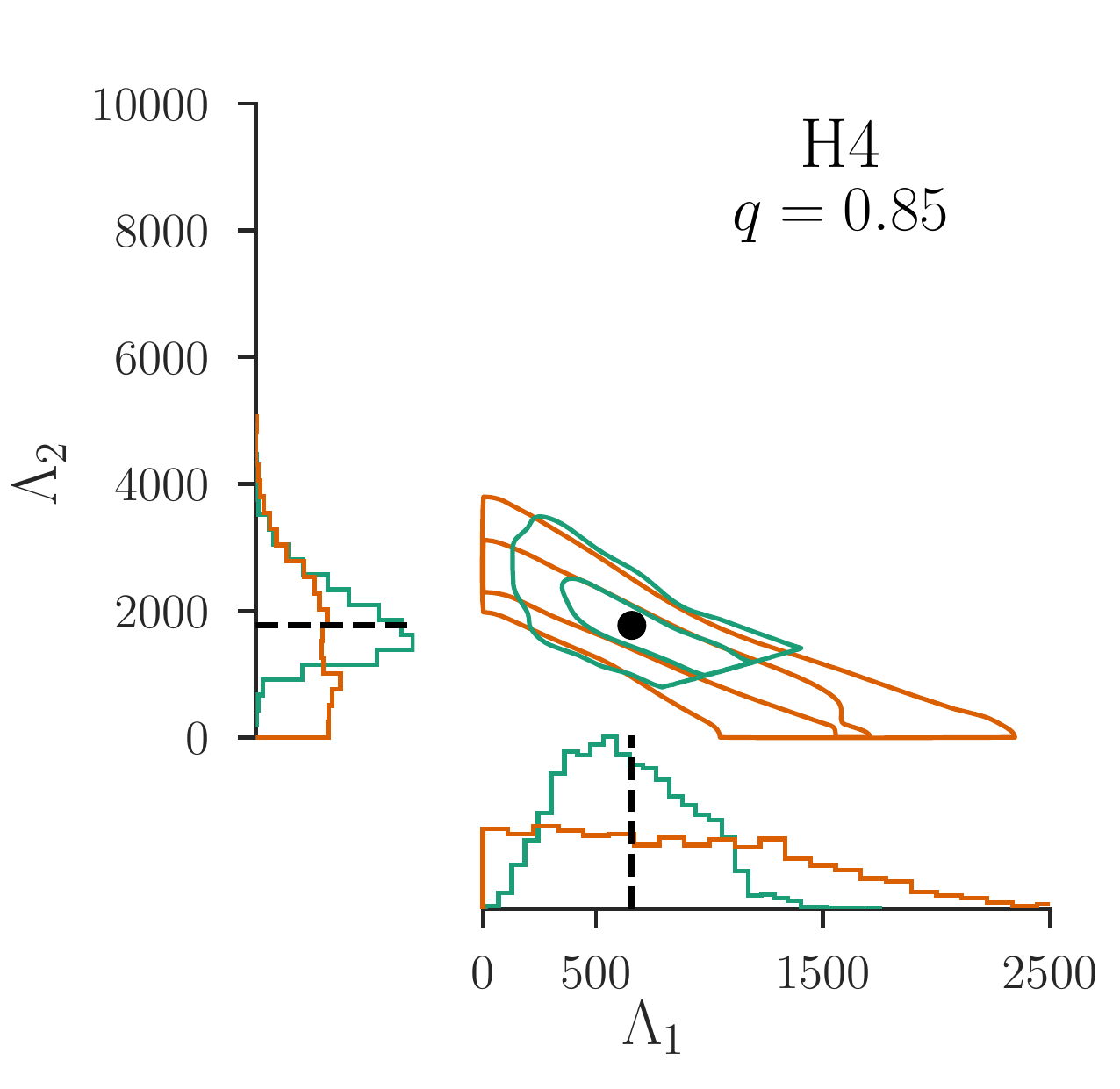}
\includegraphics[width=0.5\columnwidth,clip=true]{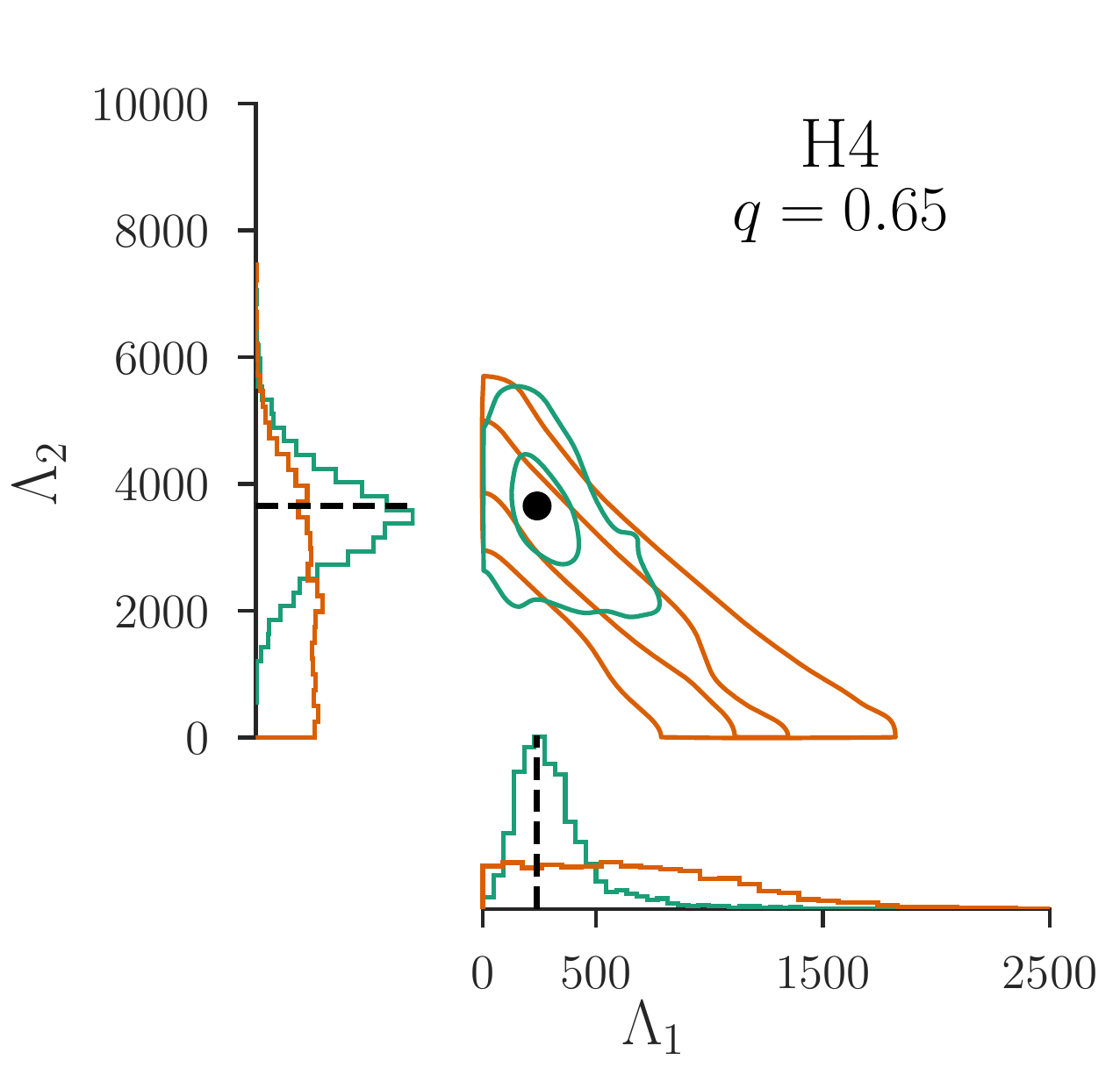}
\includegraphics[width=0.5\columnwidth,clip=true]{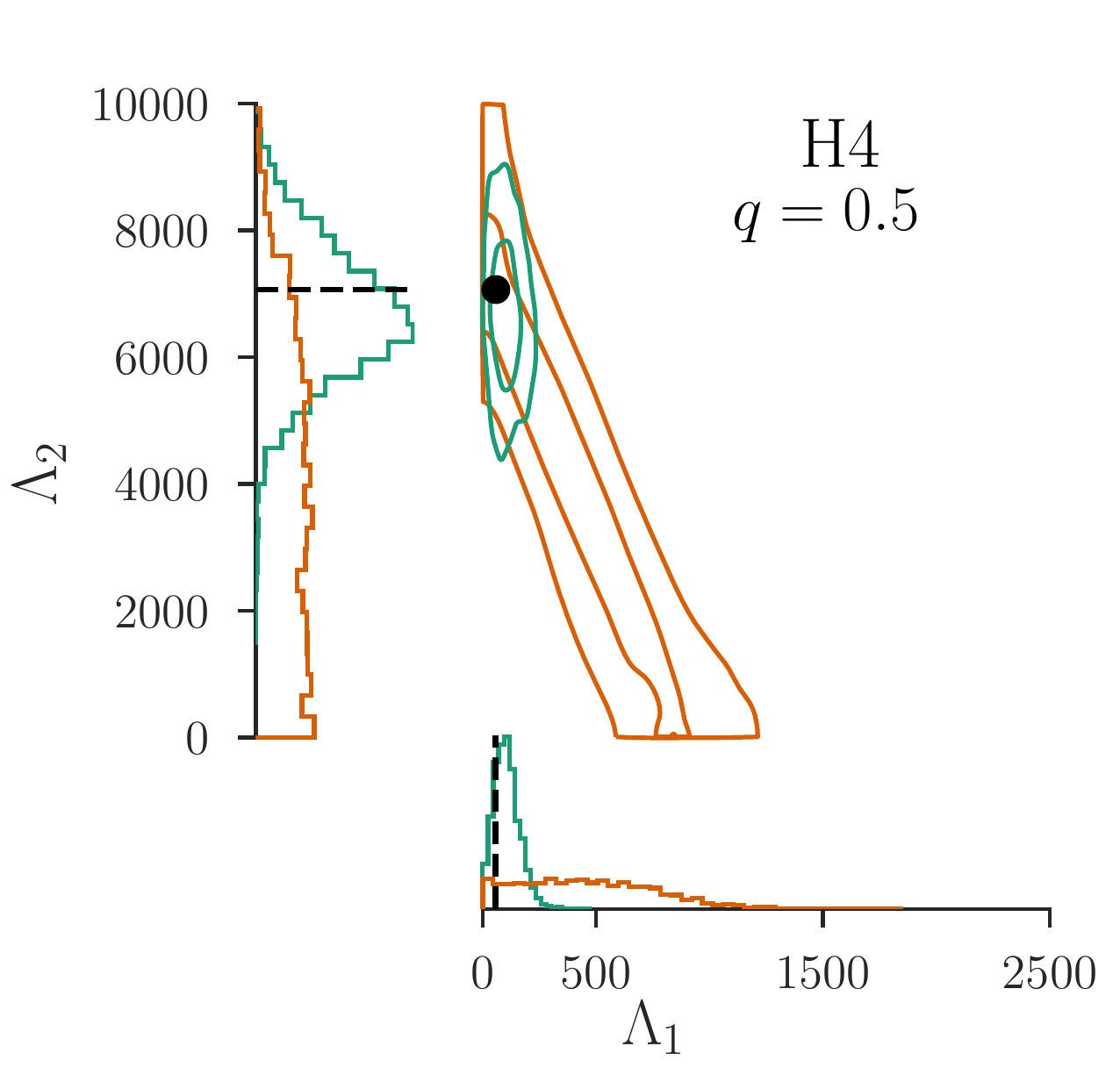}\\
\includegraphics[width=0.5\columnwidth,clip=true]{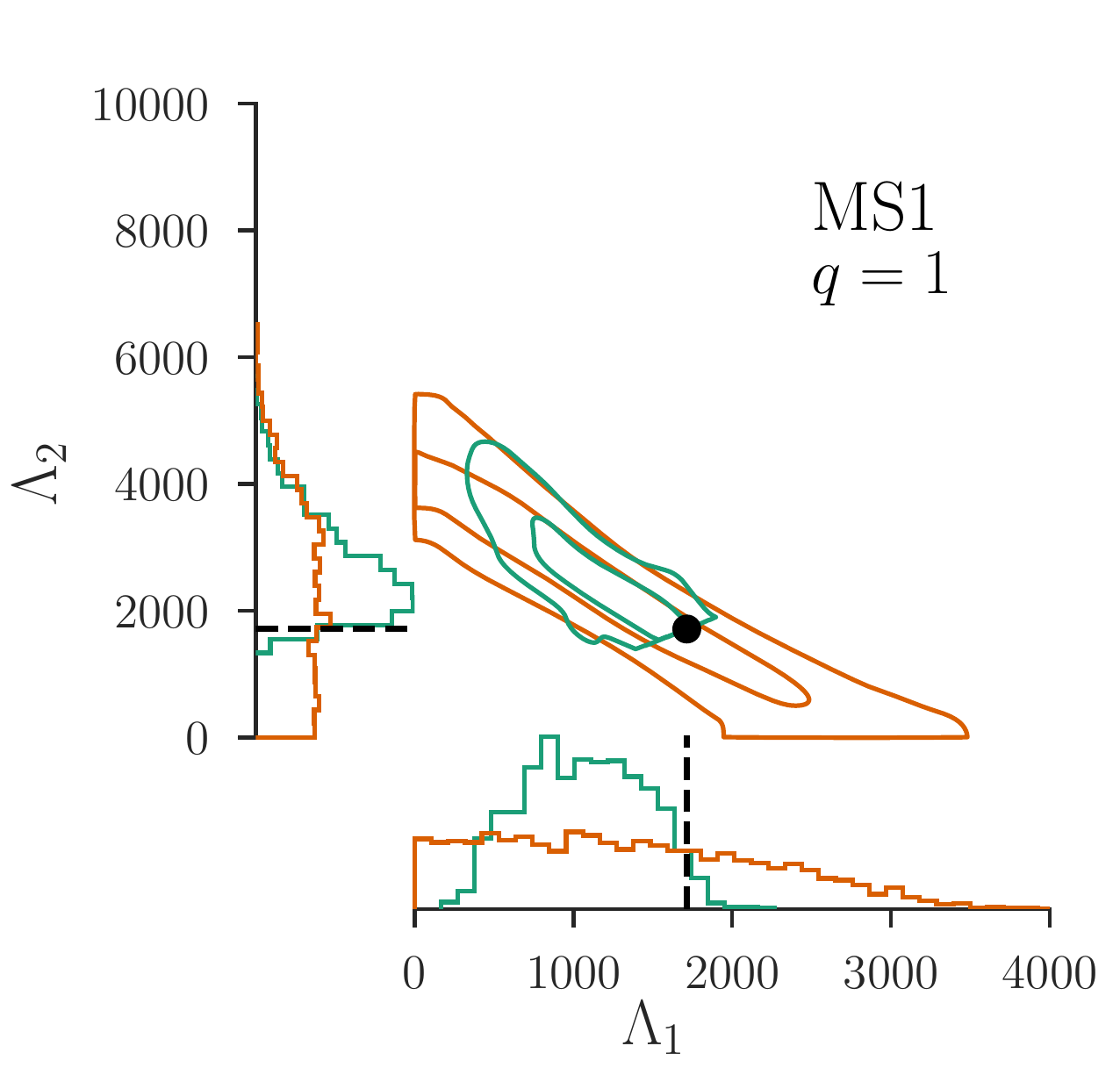}
\includegraphics[width=0.5\columnwidth,clip=true]{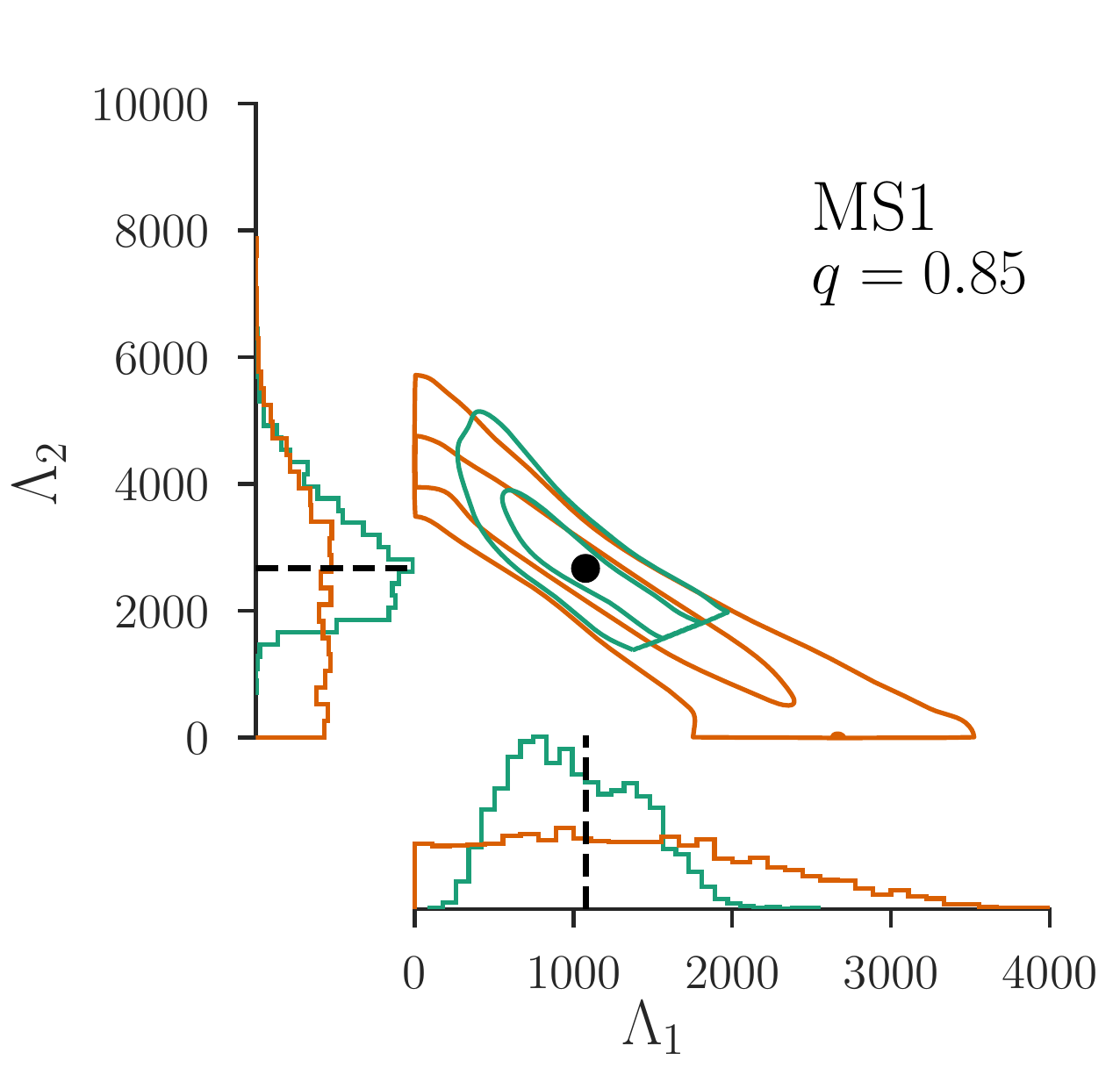}
\includegraphics[width=0.5\columnwidth,clip=true]{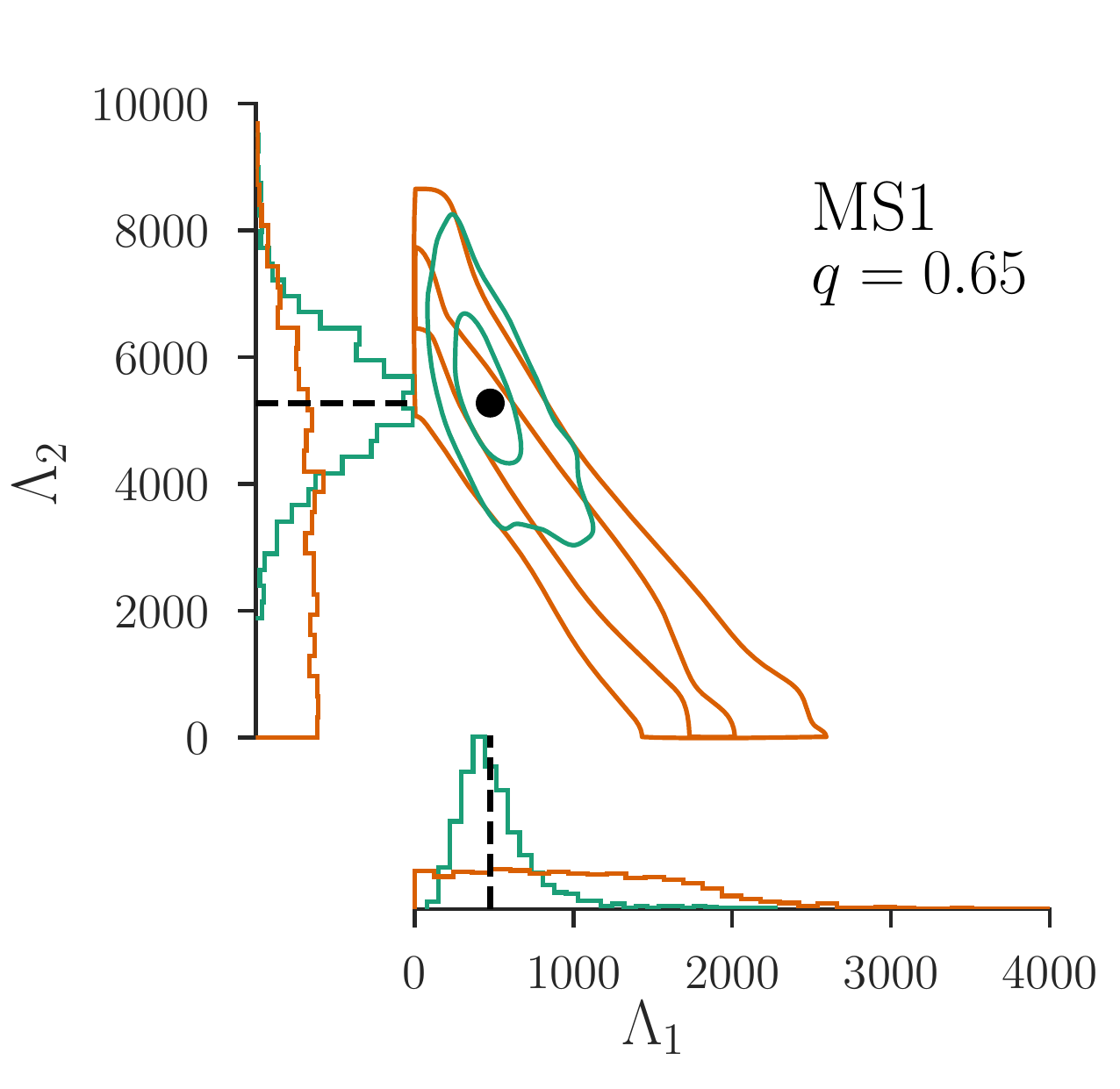}
\includegraphics[width=0.5\columnwidth,clip=true]{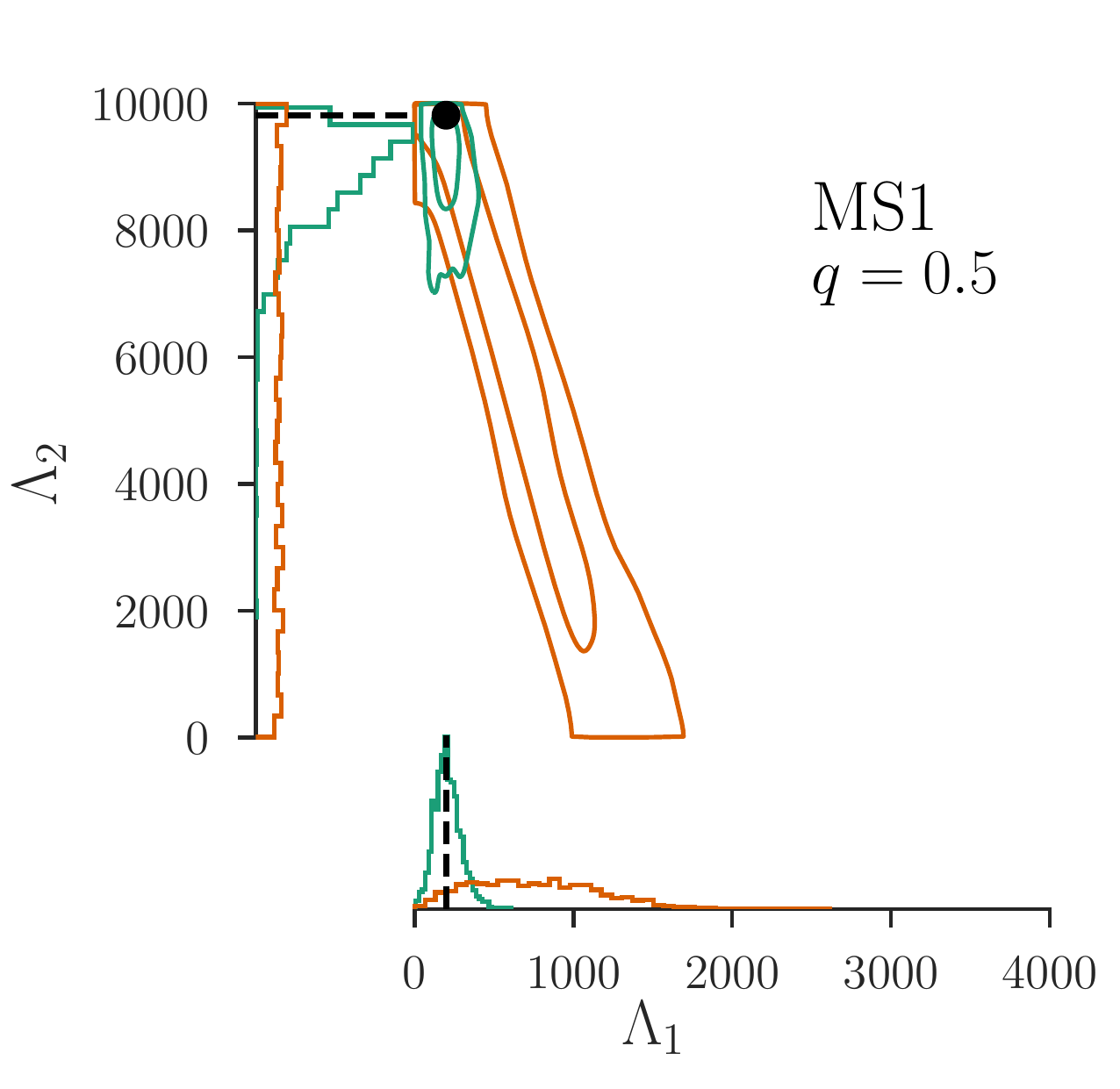}\\
\includegraphics[width=0.5\columnwidth,clip=true]{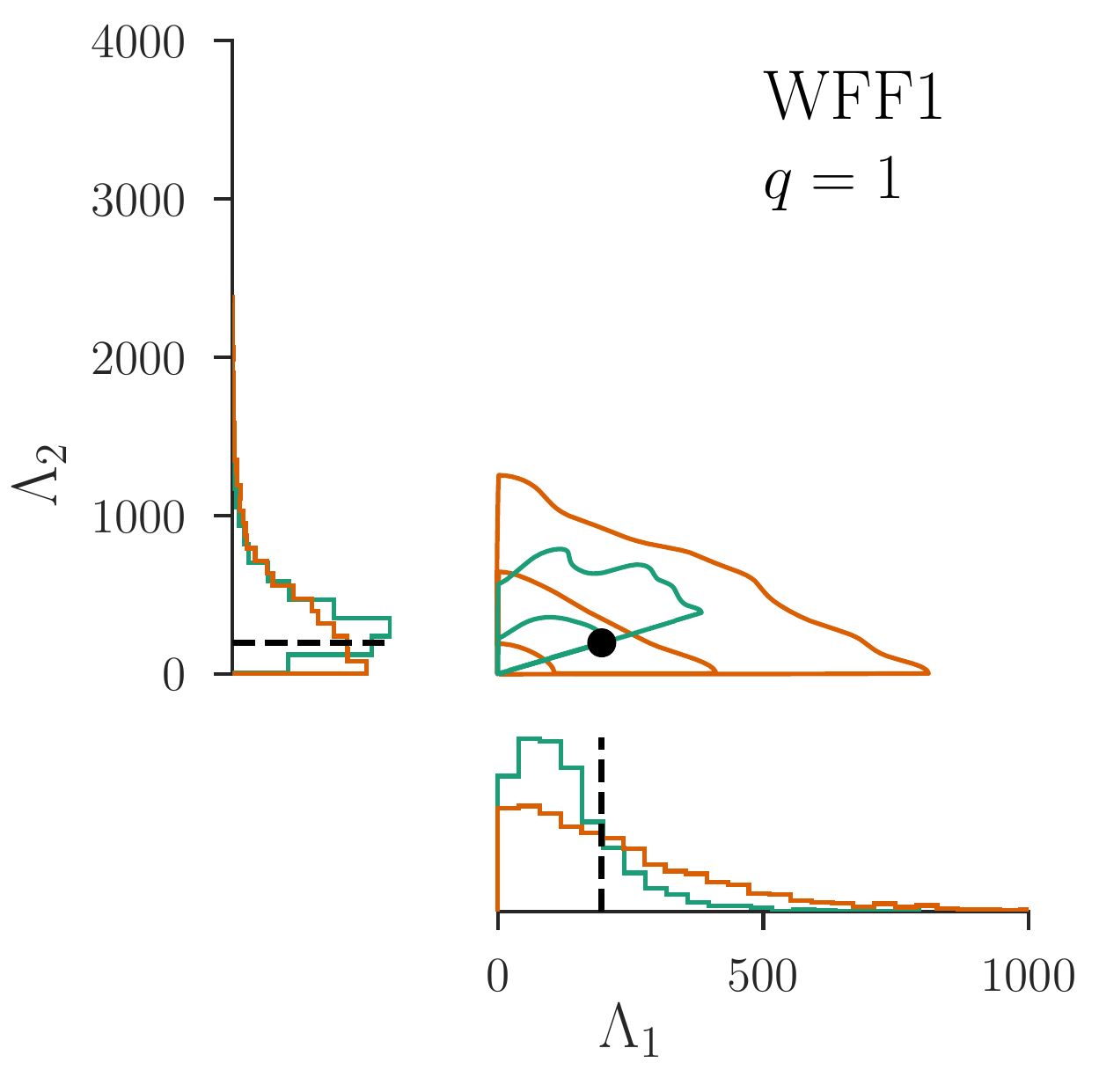}
\includegraphics[width=0.5\columnwidth,clip=true]{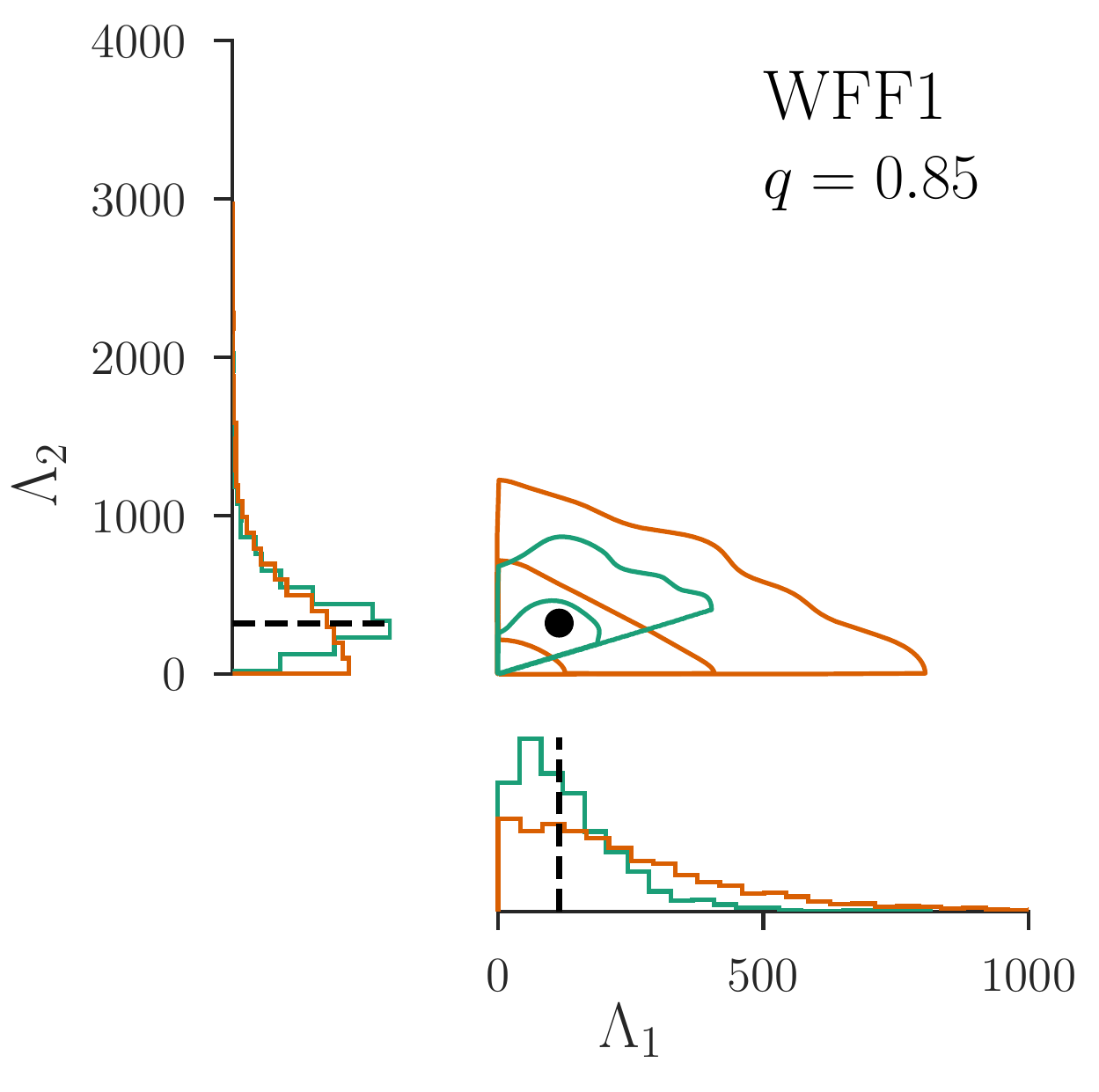}
\includegraphics[width=0.5\columnwidth,clip=true]{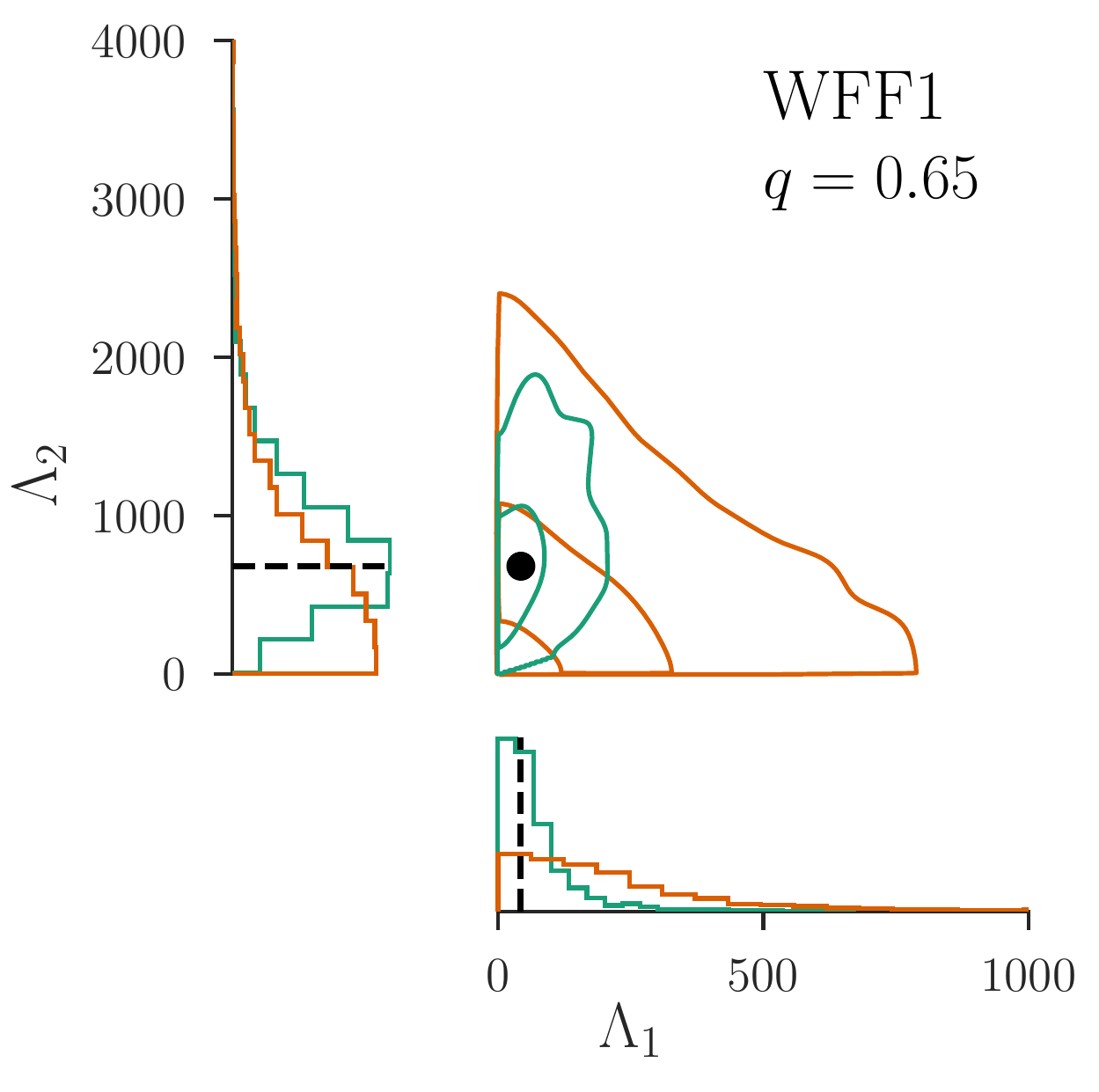}
\includegraphics[width=0.5\columnwidth,clip=true]{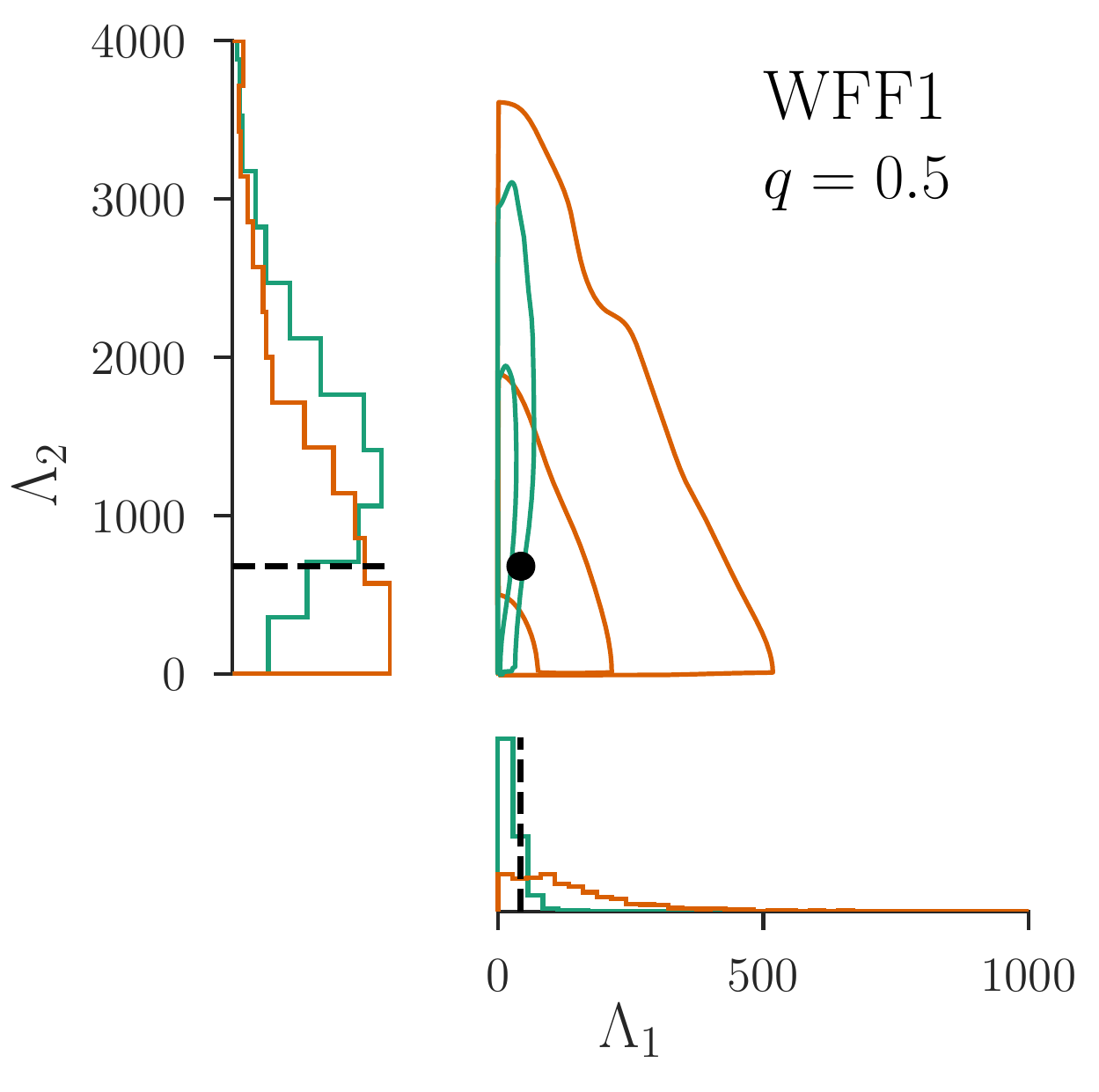}
\caption{ \label{fig:l1-l2_H4_1p0_SNR30} Two-dimensional and one-dimensional marginalized posterior density function for the dimensionless tidal deformabilities of the two stars for signals of SNR 30, the H4 (top row), MS1 (middle row), and WFF1 (bottom row) EoSs and $q=1$ (first column), $q=0.85$ (second column), $q=0.65$ (third column), and $q=0.5$ (fourth column). 
The black dot and black dashed lines indicate the injected values in each plot.  Each signal is analyzed with (green lines) and without  (orange lines) the EoS-independent relation between $\Lambda_a$ and $\Lambda_s$ and the contours shown represent the $50\%$ and $90\%$ credible regions. In all cases the EoS-independent relation leads to better measurement of the tidal parameters.}
\end{figure*}
\begin{figure*}
\includegraphics[width=0.5\columnwidth,clip=true]{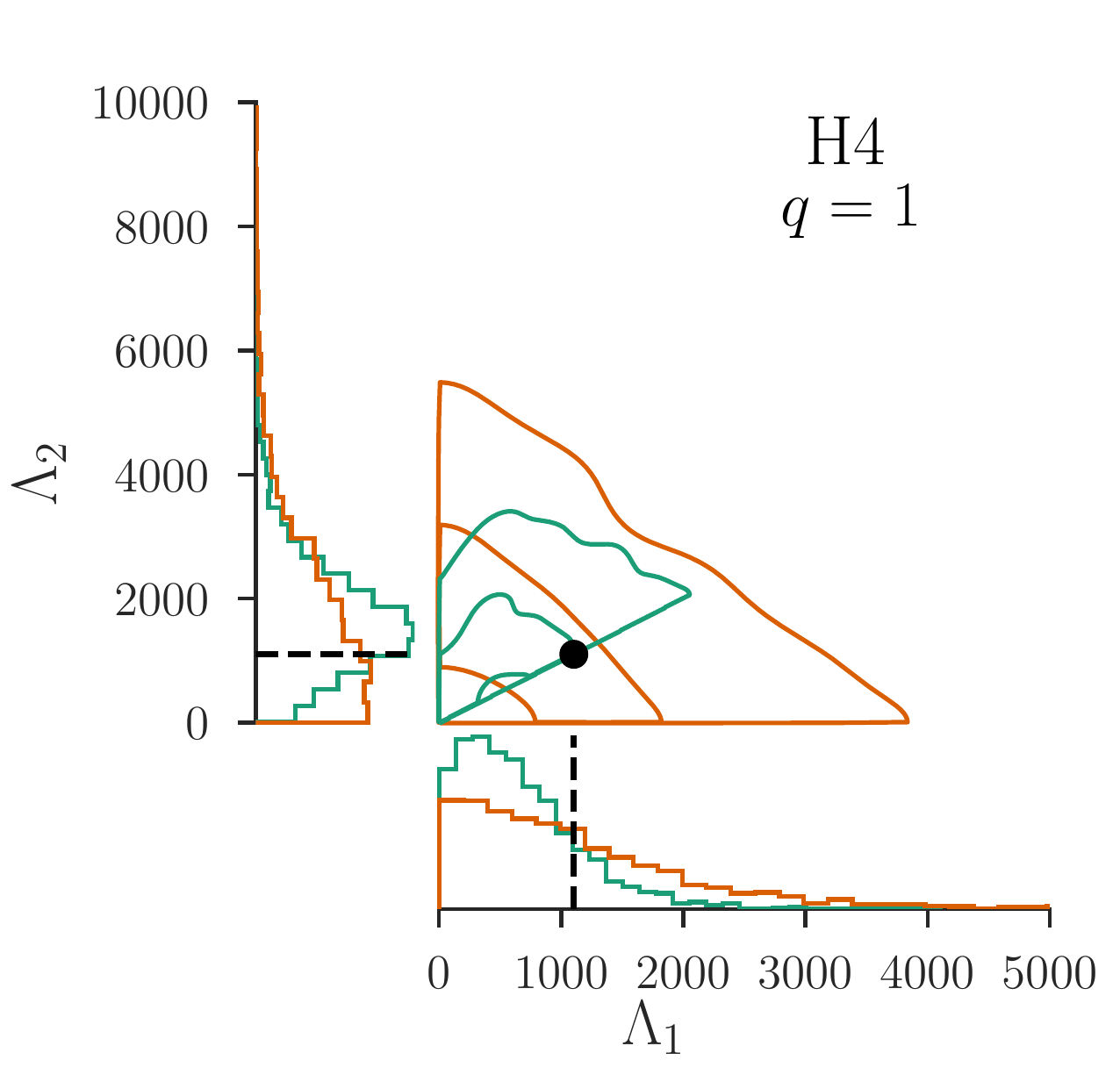}
\includegraphics[width=0.5\columnwidth,clip=true]{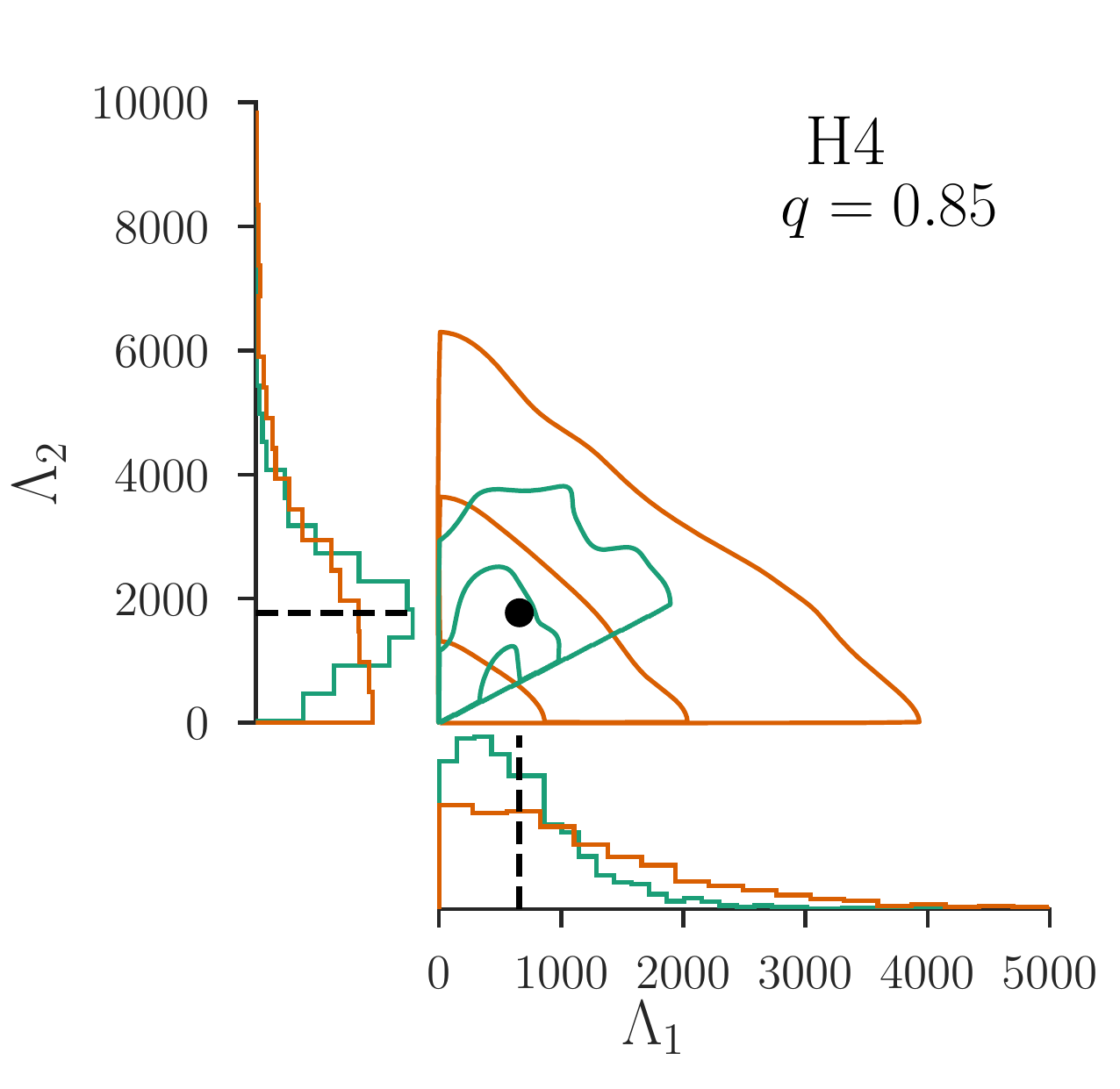}
\includegraphics[width=0.5\columnwidth,clip=true]{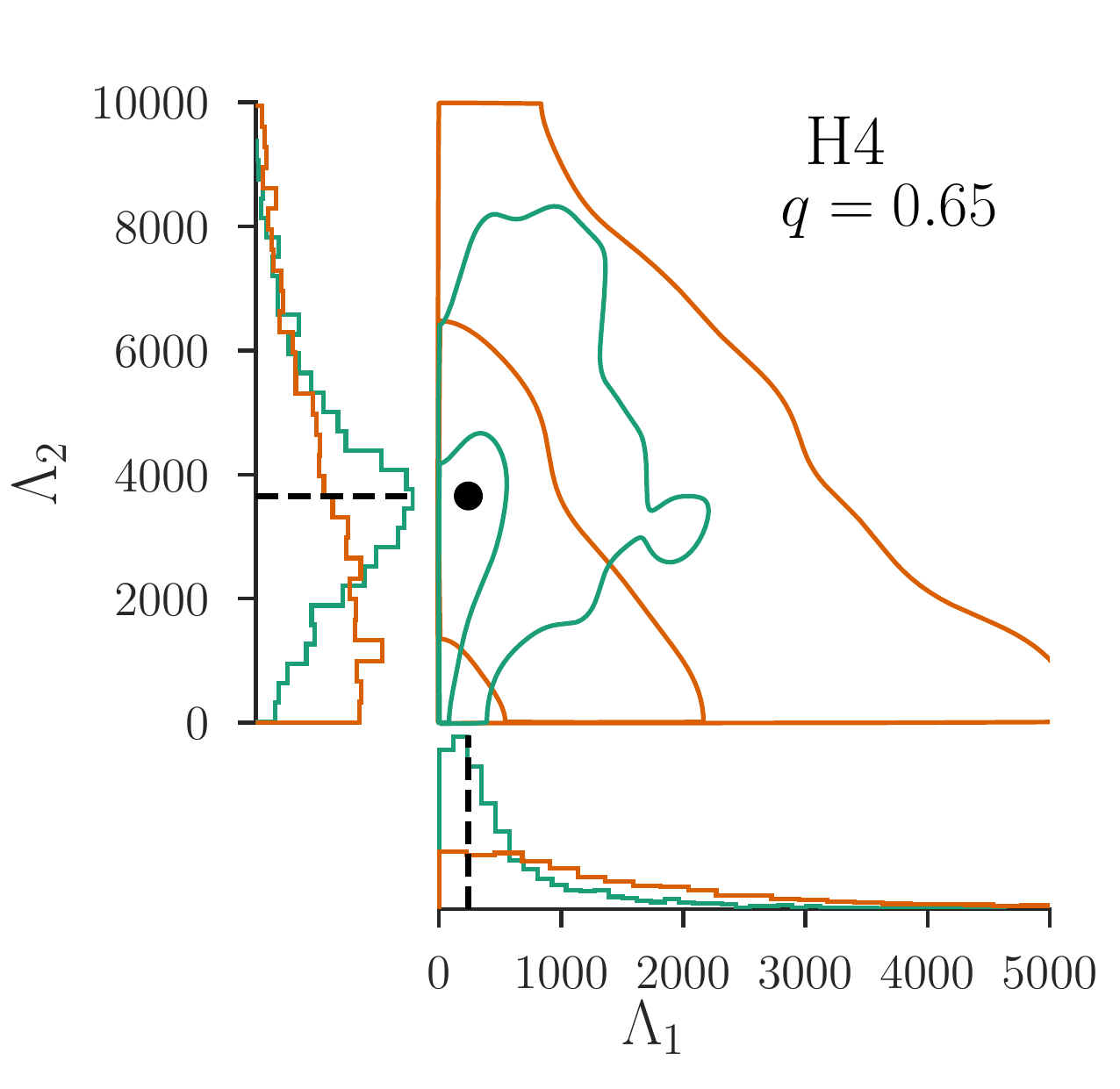}
\includegraphics[width=0.5\columnwidth,clip=true]{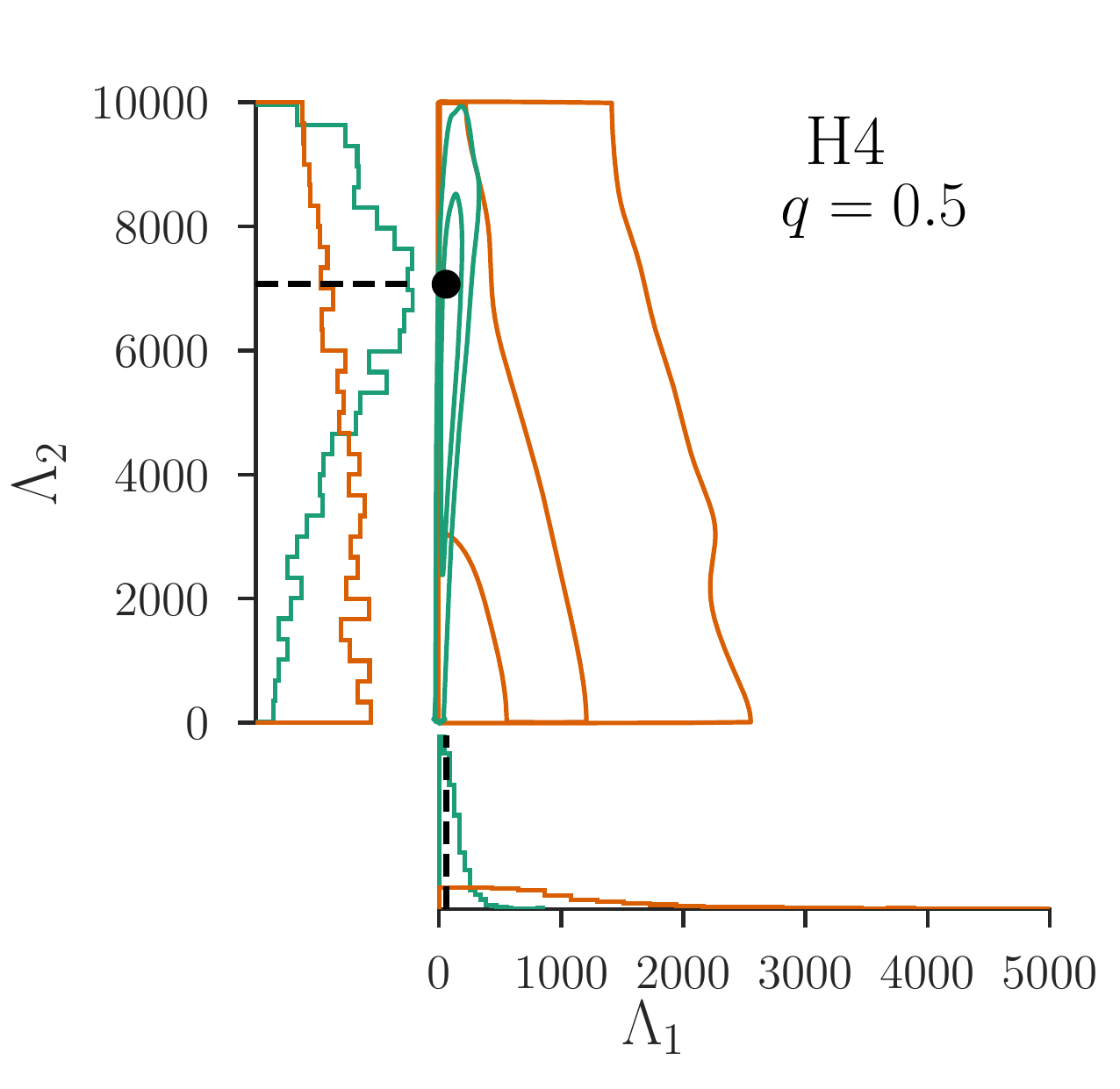}\\
\includegraphics[width=0.5\columnwidth,clip=true]{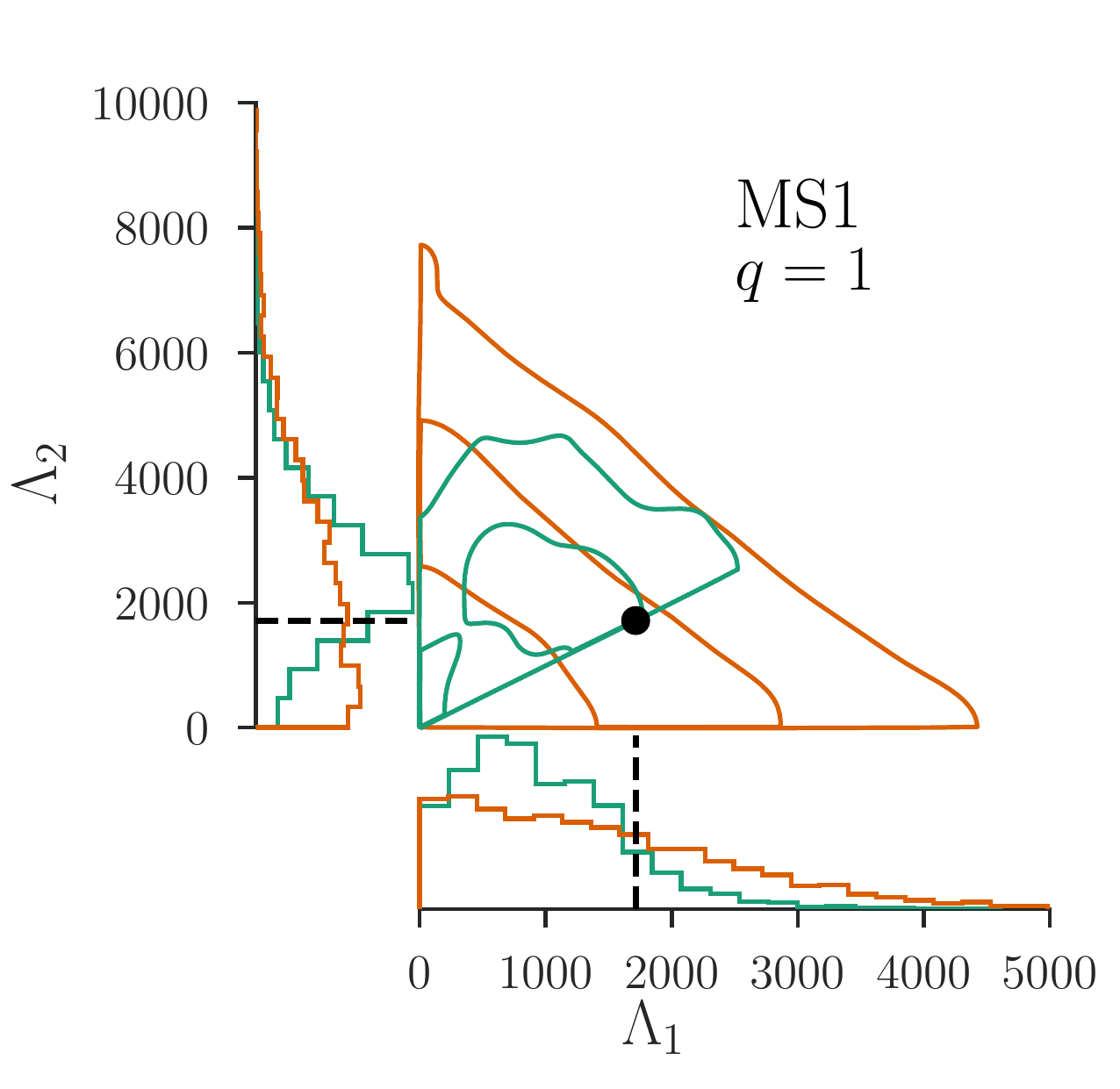}
\includegraphics[width=0.5\columnwidth,clip=true]{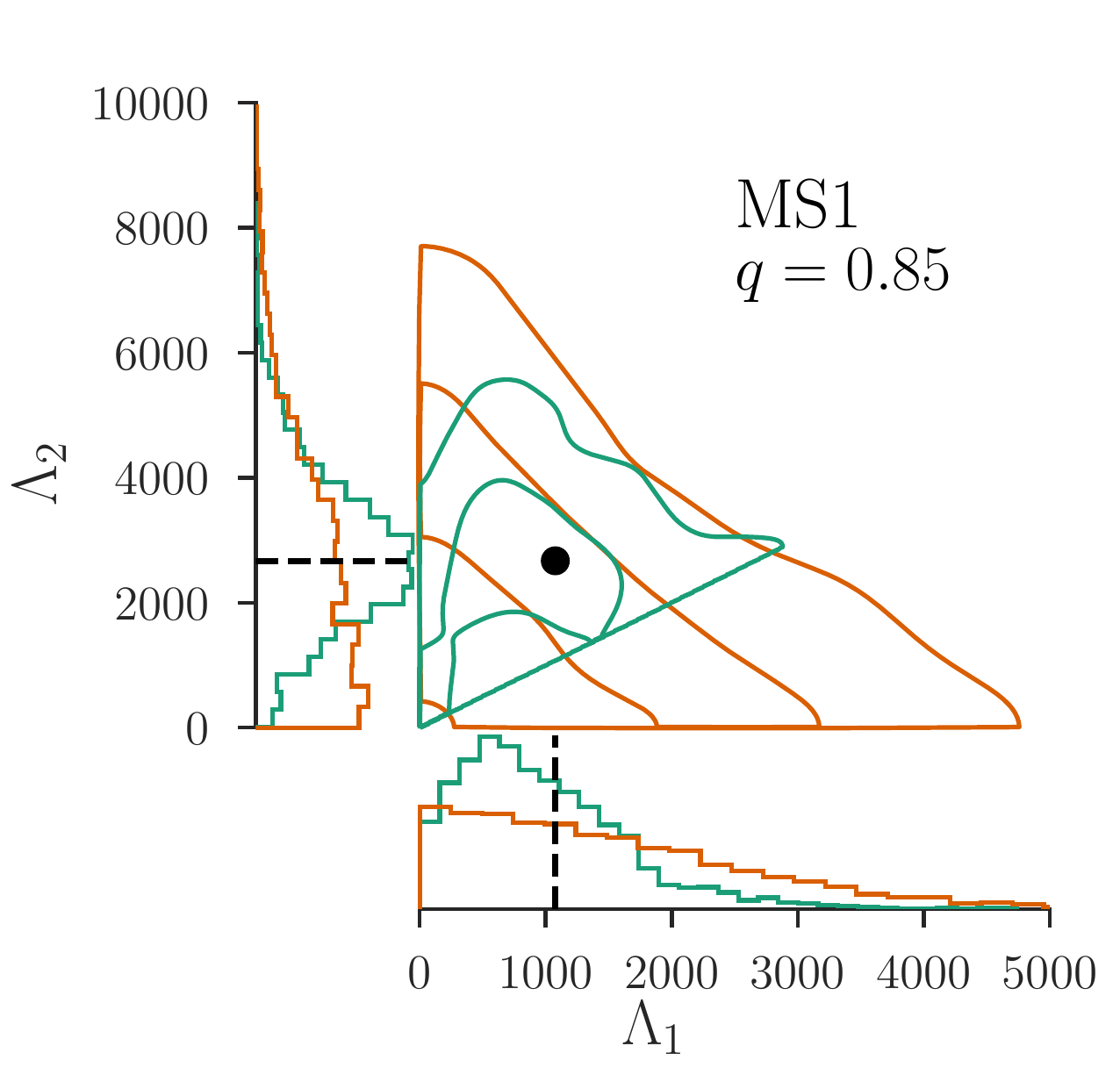}
\includegraphics[width=0.5\columnwidth,clip=true]{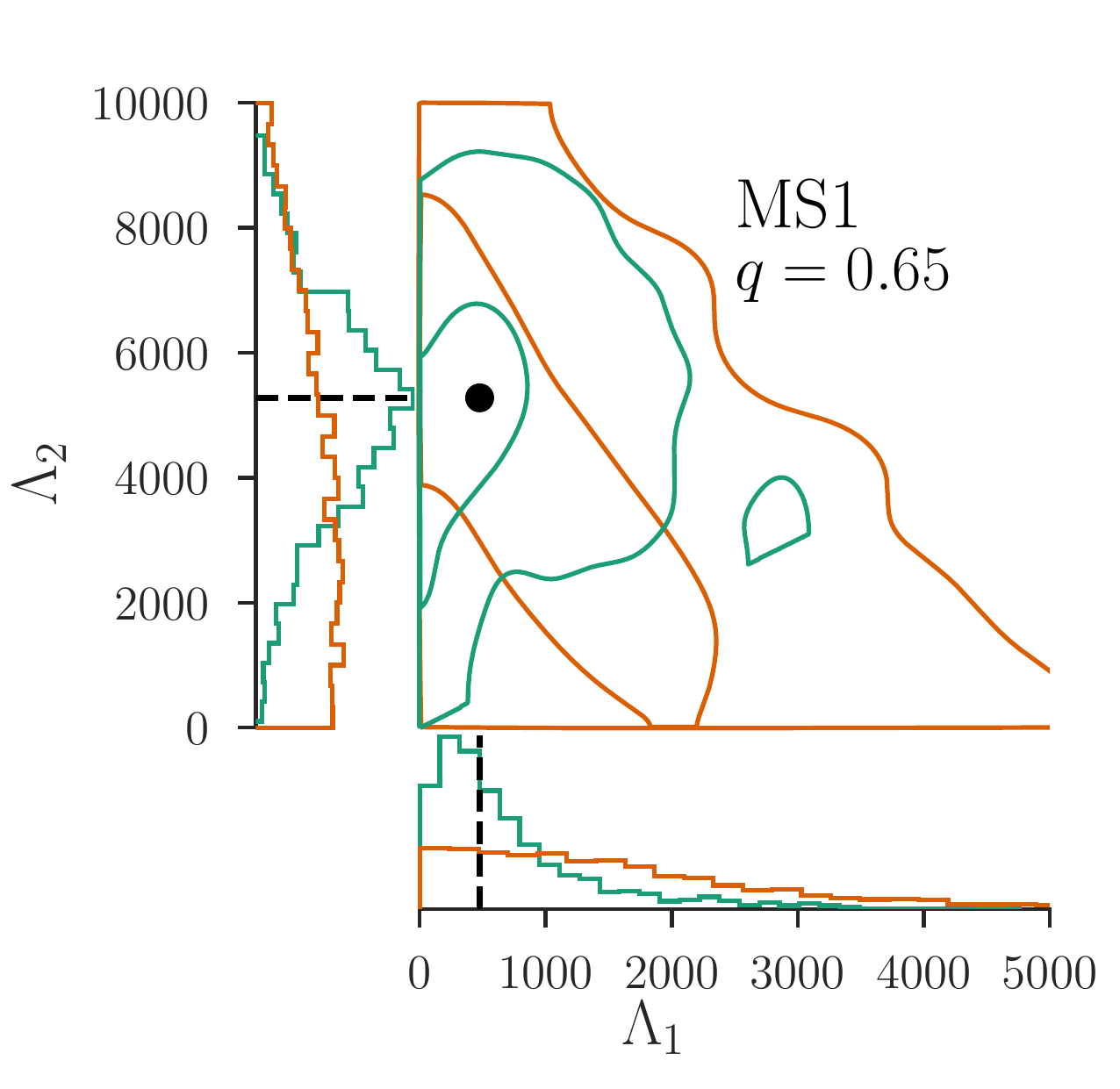}
\includegraphics[width=0.5\columnwidth,clip=true]{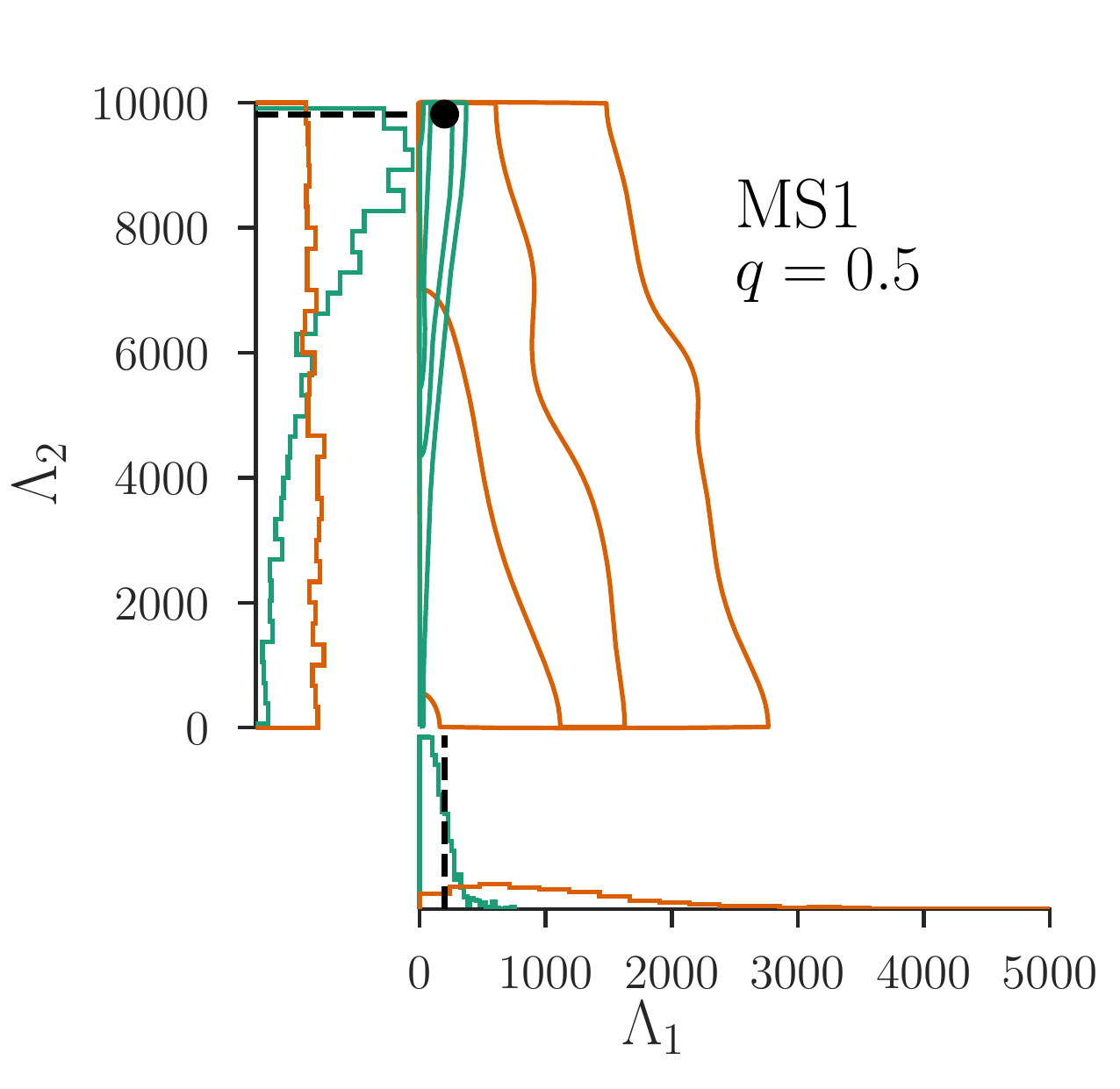}\\
\includegraphics[width=0.5\columnwidth,clip=true]{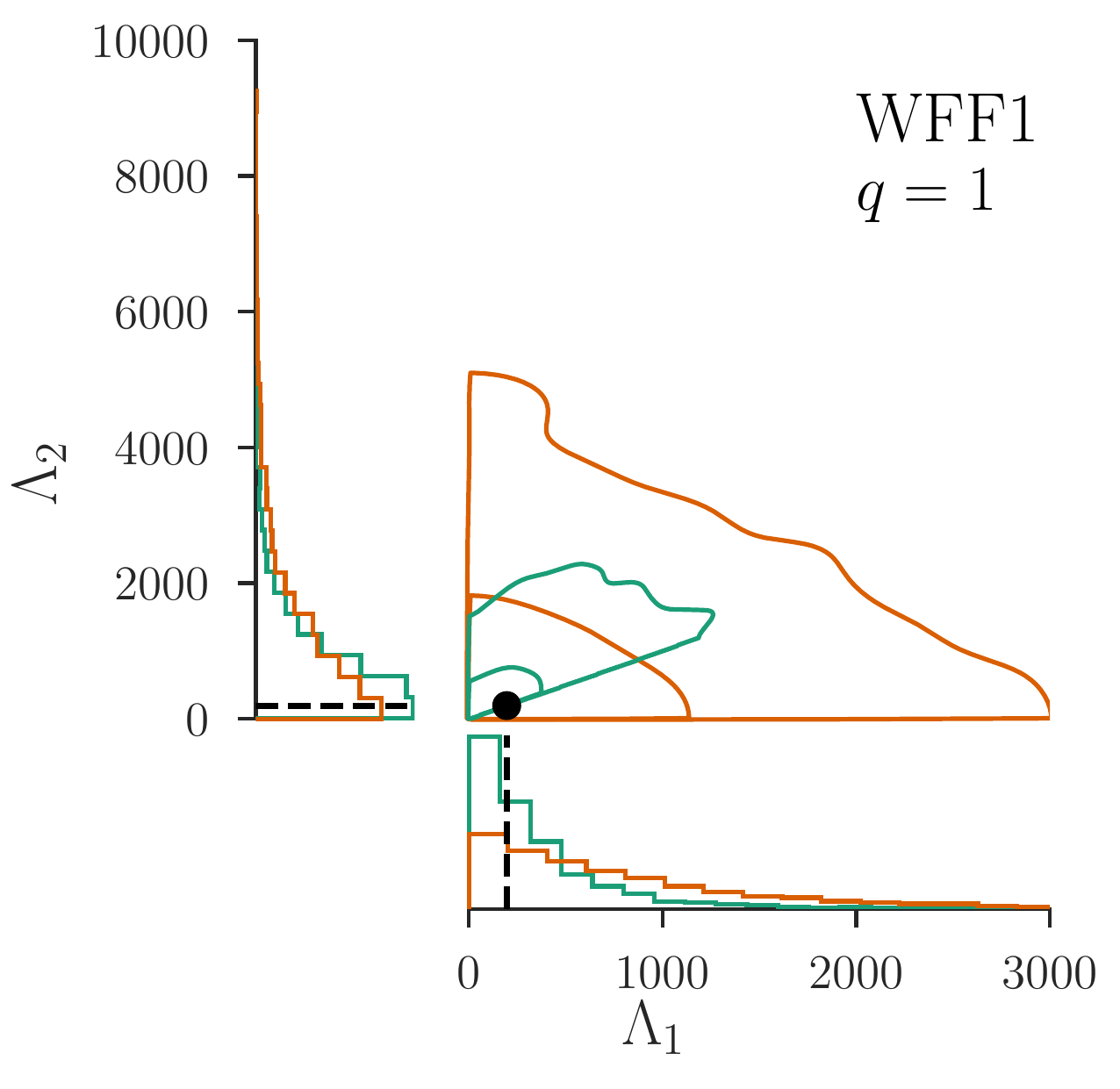}
\includegraphics[width=0.5\columnwidth,clip=true]{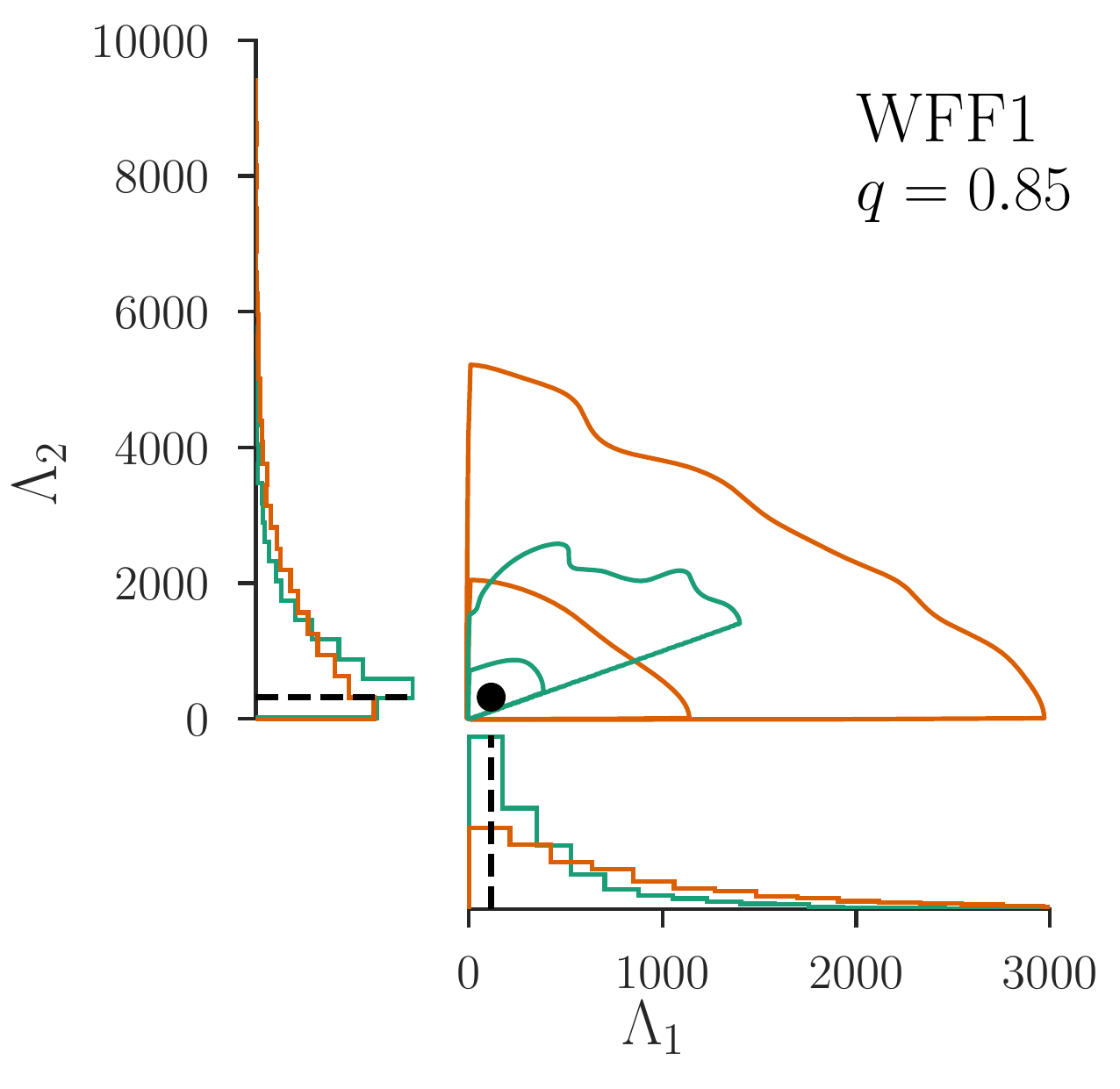}
\includegraphics[width=0.5\columnwidth,clip=true]{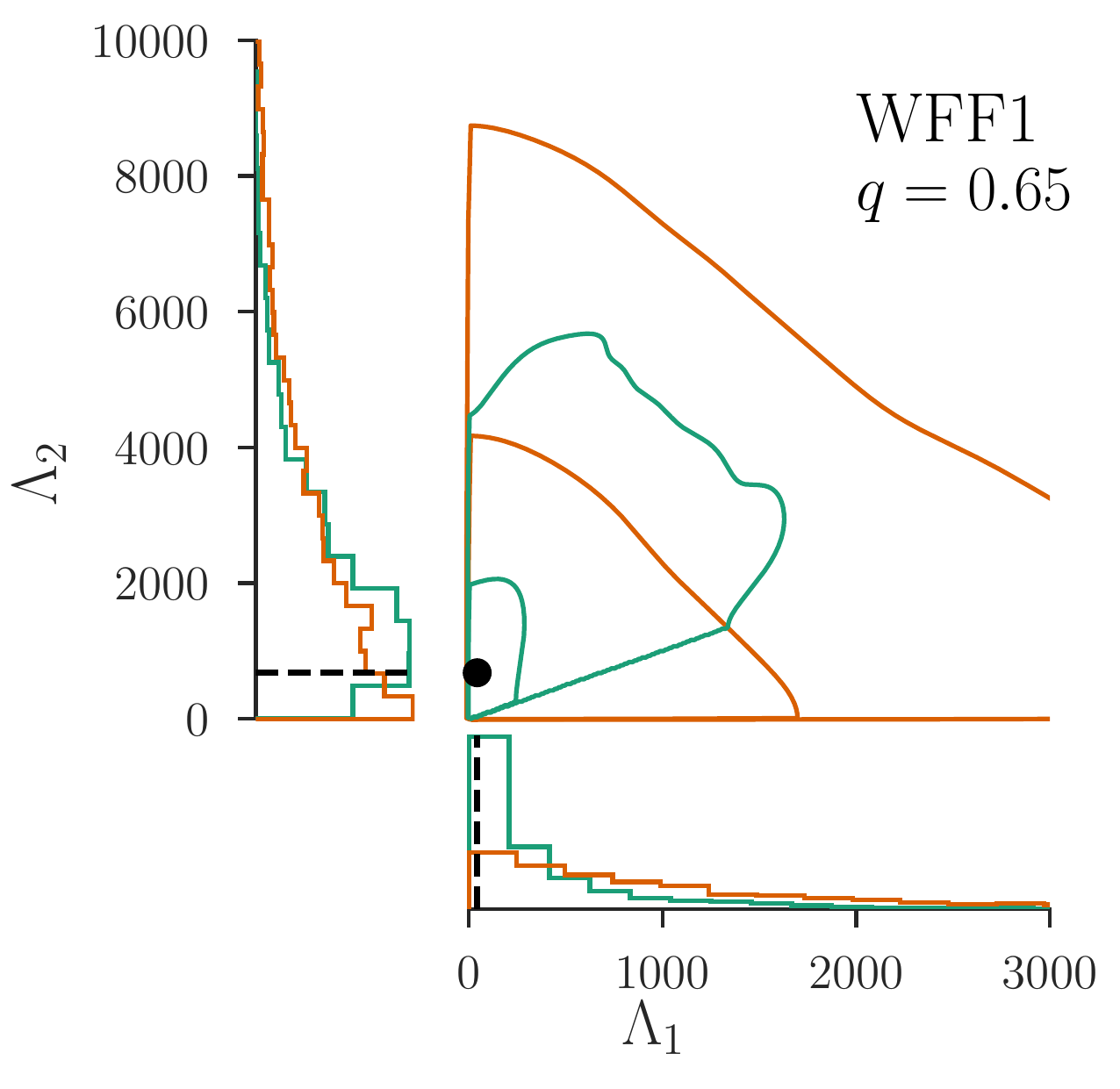}
\includegraphics[width=0.5\columnwidth,clip=true]{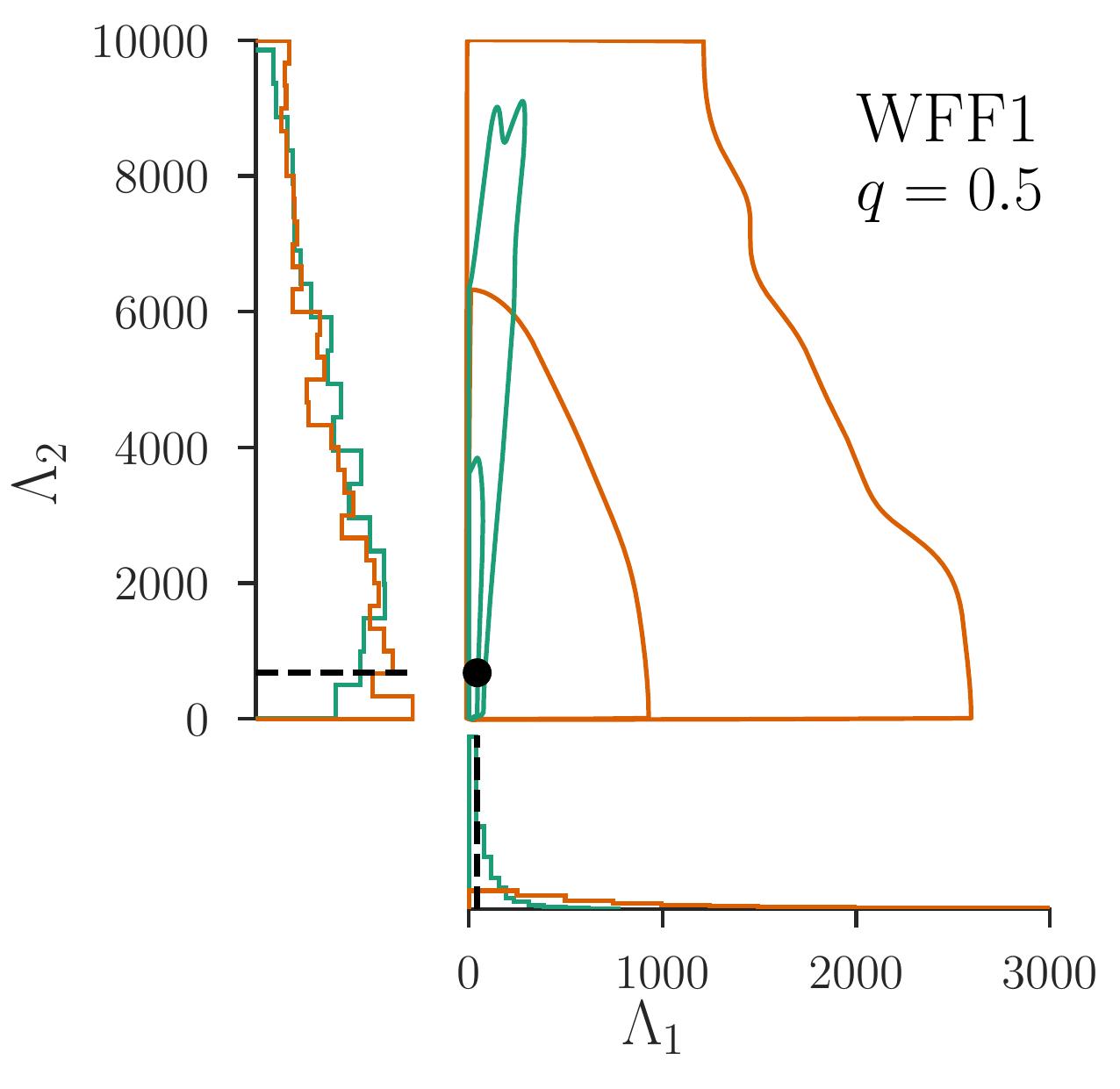}
\caption{ \label{fig:l1-l2_H4_1p0_SNR15} Similar to Fig.~\ref{fig:l1-l2_H4_1p0_SNR30} but for simulated signals with SNR 15.}
\end{figure*}

Figure~\ref{fig:l1-l2_H4_1p0_SNR30} shows the marginalized posterior density function for the tidal deformabilities of the two NSs for signals created with each EoSs, an SNR of 30, and various mass ratios. In orange we show the resulting posterior from an analysis that assumes that the tidal deformabilities and the EoSs of the two coalescing NSs are independent, while the results in green stem from the analysis that incorporates the EoS-independent relation $\Lambda_a=\Lambda_a(\Lambda_s,q;\vec{b})$, while marginalizing over the error in said relation. 

In all cases, the injected, true values (black dots for the two-dimensional plots, dashed lines for the one-dimensional posterior) fall within the posteriors calculated under the EoS-independent relation. This suggests that the error marginalization described in Sec.~\ref{sec:marg} is effective in removing any systematic biases in the results. Note that the true values usually do not fall exactly on the peak of their marginalized posterior due to parameter correlations, especially in the cases where the injected value is at the edge of the prior range ($q=1$ cases). 

Unlike the orange contours, the green posteriors do not extend into the unphysical $\Lambda_1>\Lambda_2$ region. Moreover, the green contours are less extended than the orange ones even in the $\Lambda_2>\Lambda_1$ region, overall resulting in a tighter measurement of the individual tidal deformabilities.  
For example, in the $q=1$ case with H4 (top left panel) the use of the EoS-independent relation between $\Lambda_a$ and $\Lambda_s$ reduces the measurement uncertainty of $\Lambda_1$ and $\Lambda_2$ by about a half. As the mass ratio of the system decreases, this improvement becomes more dramatic, reaching an order-of-magnitude improvement in the measurement accuracy for a mass ratio of 0.5 (top right panel). 

Besides better measurement of the tidal parameters, the EoS-independent relation can be used to place better upper bounds in the case where tidal effects are too small to measure. 
The bottom row of Fig.~\ref{fig:l1-l2_H4_1p0_SNR30} shows results for the soft EoS WFF1. 
Despite not leading to an unambiguous measurement of tidal effects for all mass ratios ($\Lambda_1=\Lambda_2=0$ is consistent with this signal at the $90\%$ credible level), the EoS-independent relation leads to a tighter upper bound on the tidal deformabilities and, as a result, on the NS radius.

We arrive at qualitatively similar conclusions from our set of injections with an SNR 15, as also shown in Fig.~\ref{fig:l1-l2_H4_1p0_SNR15}. In all cases the EoS-independent relation leads to a better measurement of the tidal parameters, though the absolute magnitude of the credible regions and credible intervals is larger owing to the reduced signal strength.

\subsection{Distinguishing between BNSs and NSBHs}

The middle row of Fig.~\ref{fig:l1-l2_H4_1p0_SNR30} shows results for the MS1 EoS, a stiff EoS that was shown to be inconsistent with GW170817~\cite{TheLIGOScientific:2017qsa}. In the context of this study, therefore, we use it only for demonstration and find that the larger the NS radius, the easier it is to detect tidal effects in the observed GW.  
Examining the results for MS1, it is apparent that despite the large SNR of the signal and the large tidal effects present in the waveform, the orange contours are consistent with an NSBH system, i.e.~at least one of the two binary components is consistent with $\Lambda_i=0$ at the $50\%$ level. 
This means that it will be hard to distinguish between BNS and NSBH systems with GWs alone if we assume that $\Lambda_1$ and $\Lambda_2$ are independent~\cite{Yang:2017gfb}. 

The use of the EoS-independent relation offers an opportunity to separate BNS and NSBH systems on the basis of tidal measurements alone. 
In the context of Bayesian Inference, selection between competing models for the data -- in this case the BNS and the NSBH models -- amounts to comparing the probability that each model is correct given common data. 
Assuming equal prior odds for each model, the ratio of the probabilities is the ratio of the evidence for each model, termed the \emph{Bayes Factor}. 
While we do not study NSBH systems here, we have computed the evidence $\mathcal{Z}$ for the BNS model assuming either independent tides, $\mathcal{Z}_{\rm{ind}}$, or by imposing the EoS-independent relation, $\mathcal{Z}_{\rm{rel}}$. 
In all cases we found $\mathcal{Z}_{\rm{rel}}/\mathcal{Z}_{\rm{ind}}\sim\mathcal{O}(10)$. 
This suggests that our analysis will lead to larger Bayes Factors in favor of the BNS interpretation of these signals, making it easier to distinguishing them from NSBHs. 
We leave a detailed study of the Bayes Factors in favor of BNS or NSBH systems for different systems and EoSs for future work.

\subsection{Measurement of  $\tilde{\Lambda}$ and the mass ratio}

\begin{figure}[tb]
\includegraphics[width=\columnwidth,clip=true]{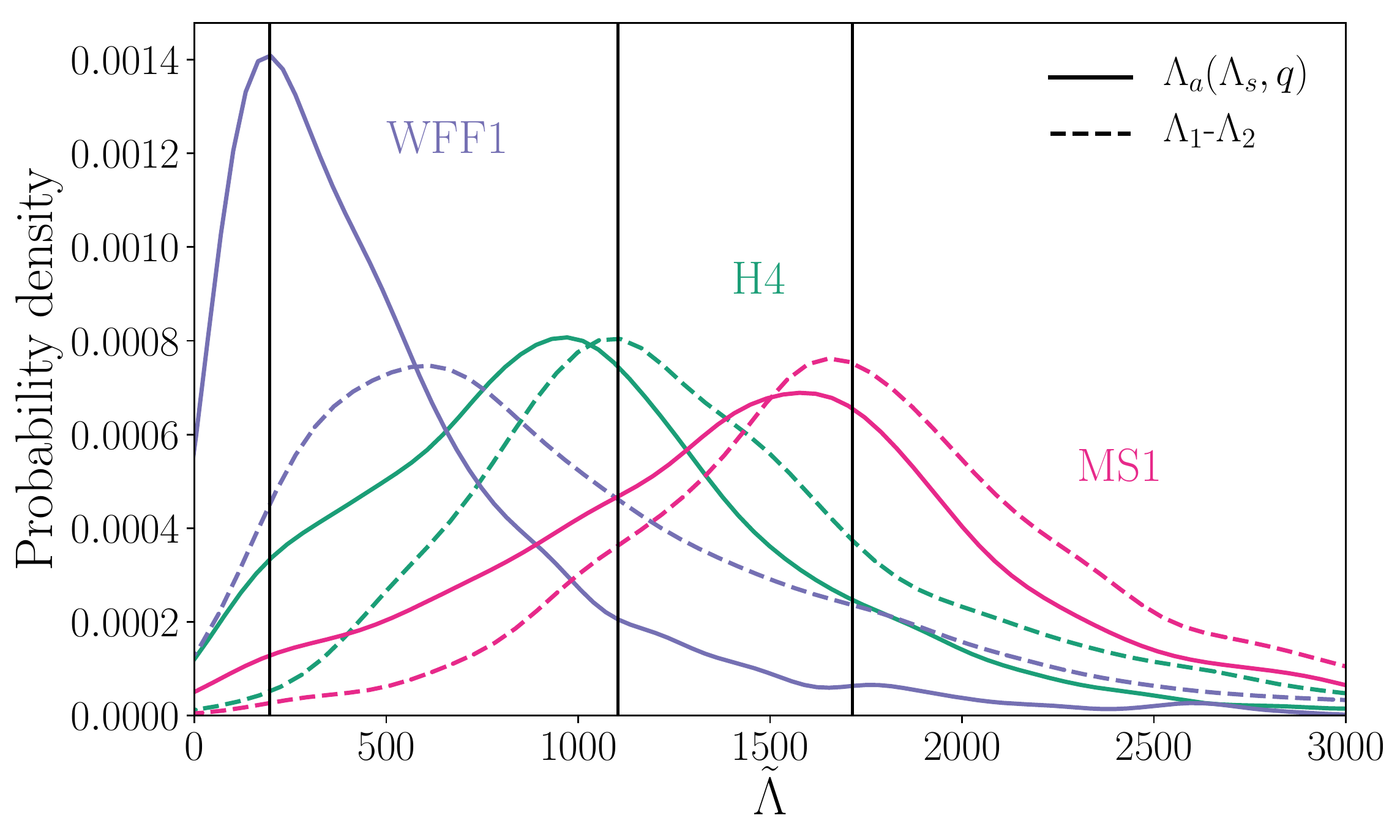}
\includegraphics[width=\columnwidth,clip=true]{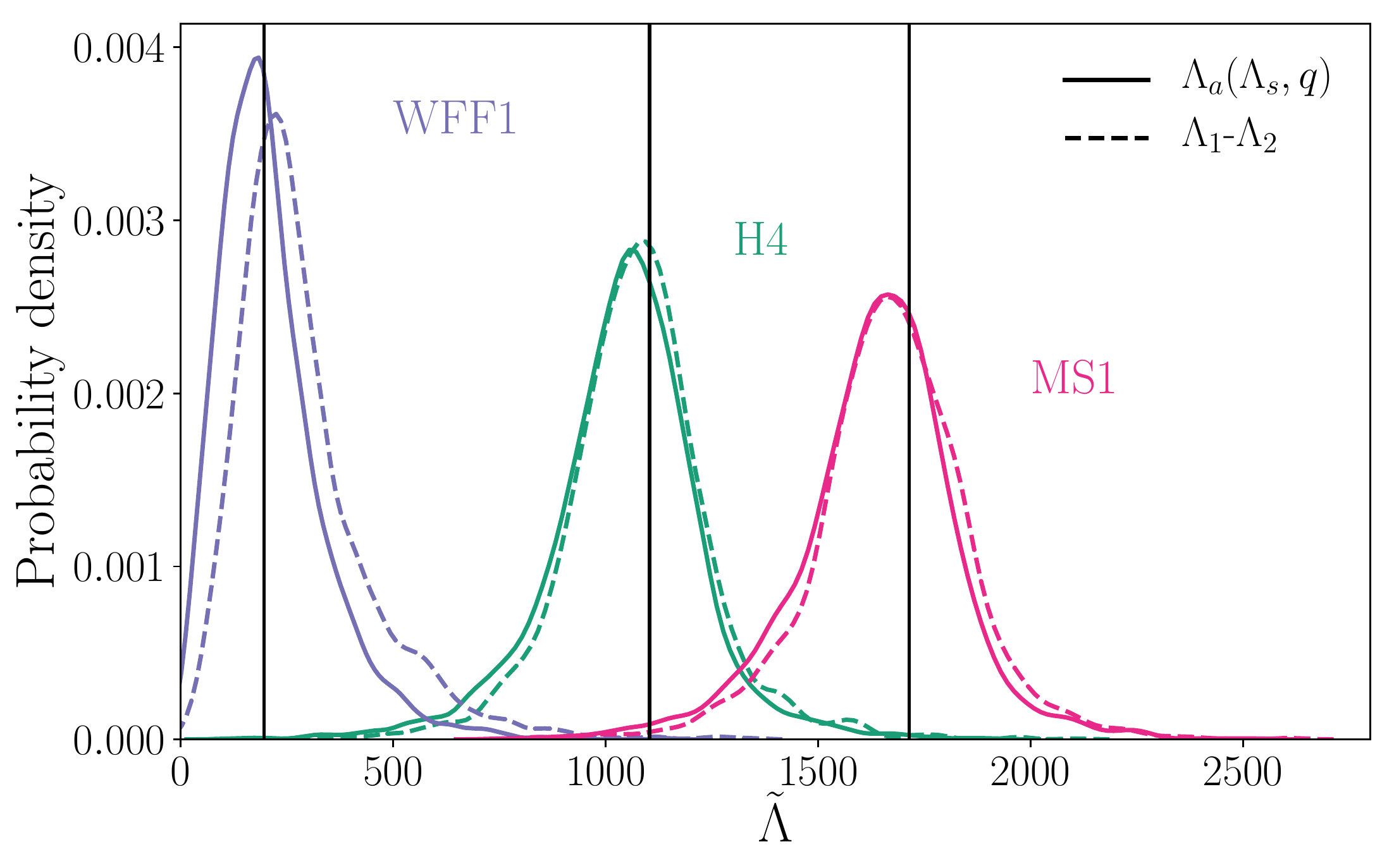}
\caption{ \label{fig:LambdaT_H4_1p0} Marginalized posterior for $\tilde{\Lambda}$ for various EoSs and SNR 15 (top panel) and 30 (bottom pannel), and a mass ratio of $1$ when assuming that the individual tidal deformabilities are independent (dashed lines), and when we employ the EoS-independent relation (solid lines). 
For low SNR there are apparent differences between the $\tilde{\Lambda}$ posteriors obtained under there two approaches. 
At higher SNR both posteriors are very similar, as expected from the fact that $\tilde{\Lambda}$ is a well-measured parameter.
}
\end{figure}

In Fig.~\ref{fig:LambdaT_H4_1p0} we show the marginalized posterior density for $\tilde{\Lambda}$, the best measured tidal parameter. We present results for signals produced by equal-mass NSs, the three EoS studied here at SNR 15 (top panel) and 30 (bottom panel). Despite $\tilde{\Lambda}$ being the best measured tidal parameter, the SNR 15 posteriors are mildly affected by the use of the EoS-independent relation.
However, for larger SNR signals the $\tilde{\Lambda}$ posterior is only very mildly affected by the EoS-independent relation, as expected from a parameter whose measurement is strongly informed by the data.

As discussed in Sec.~\ref{signals} part of the difference between the posteriors obtained with the different methods can be attributed to the fact that the two analysis assume different priors for the tidal parameters. As expected from the shape of the tidal priors in Fig.~\ref{fig:priors}, the $\tilde{\Lambda}$ posterior shifts towards lower values when we use the $\Lambda_a=\Lambda_a(\Lambda_s,q;\vec{b})$ relation and assume a BNS system. However the effect of the prior is smaller in the case of the well measured parameter $\tilde{\Lambda}$ than the individual tidal parameters in Fig.~\ref{fig:l1-l2_H4_1p0_SNR30}.

Besides the tidal parameters, the mass ratio of the system appears in the EoS-independent relation as well. 
Despite that, its measurability is not expected to be significantly affected by the use of the relation, since it first enters at 1PN order in the waveform phase, a total of 4PN orders before the tidal parameters. 
Therefore our ability to measure it should not rely on the tidal measurement. 

\begin{figure}[tb]
\includegraphics[width=0.95\columnwidth,clip=true]{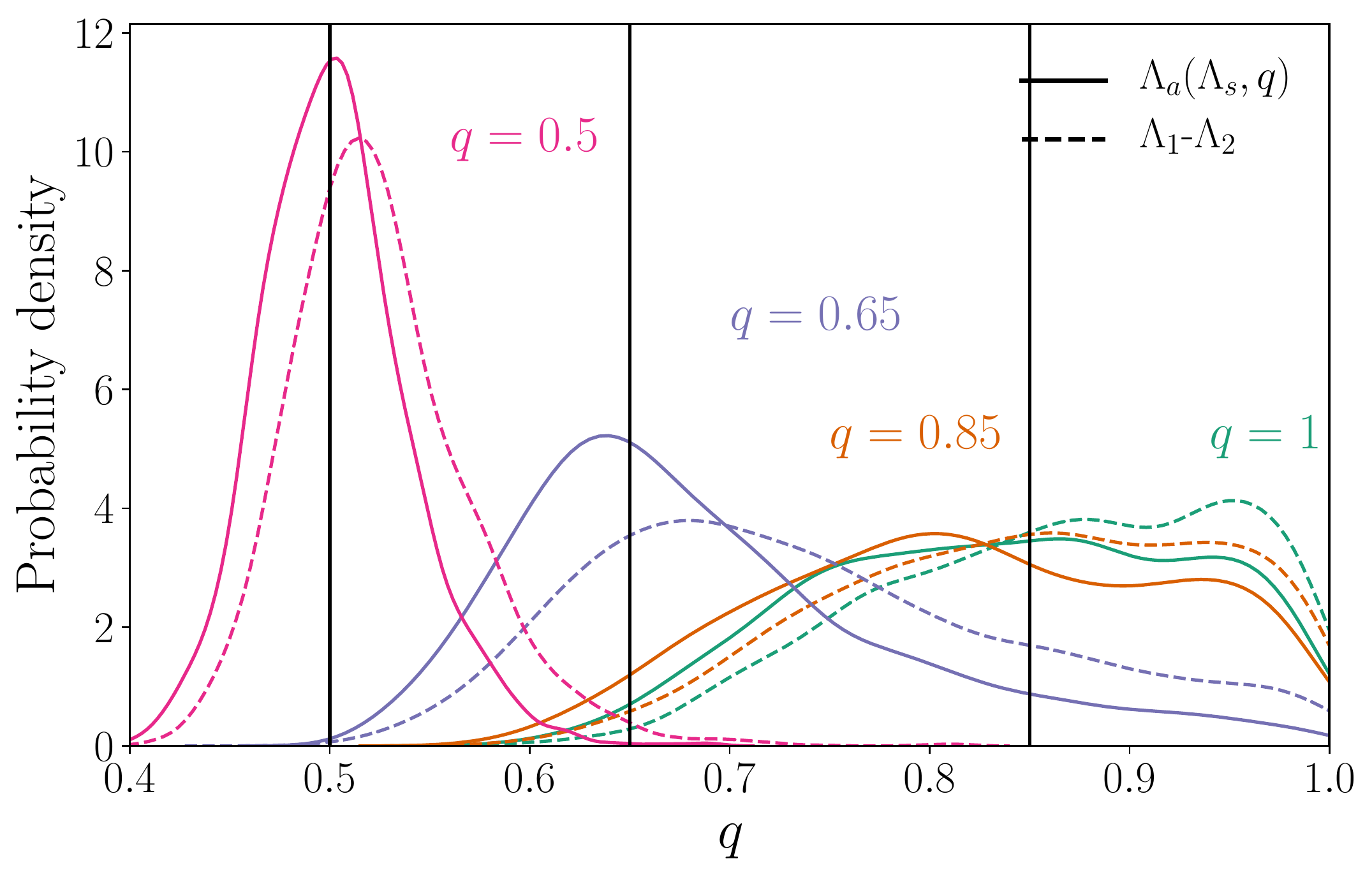}\\
\includegraphics[width=0.95\columnwidth,clip=true]{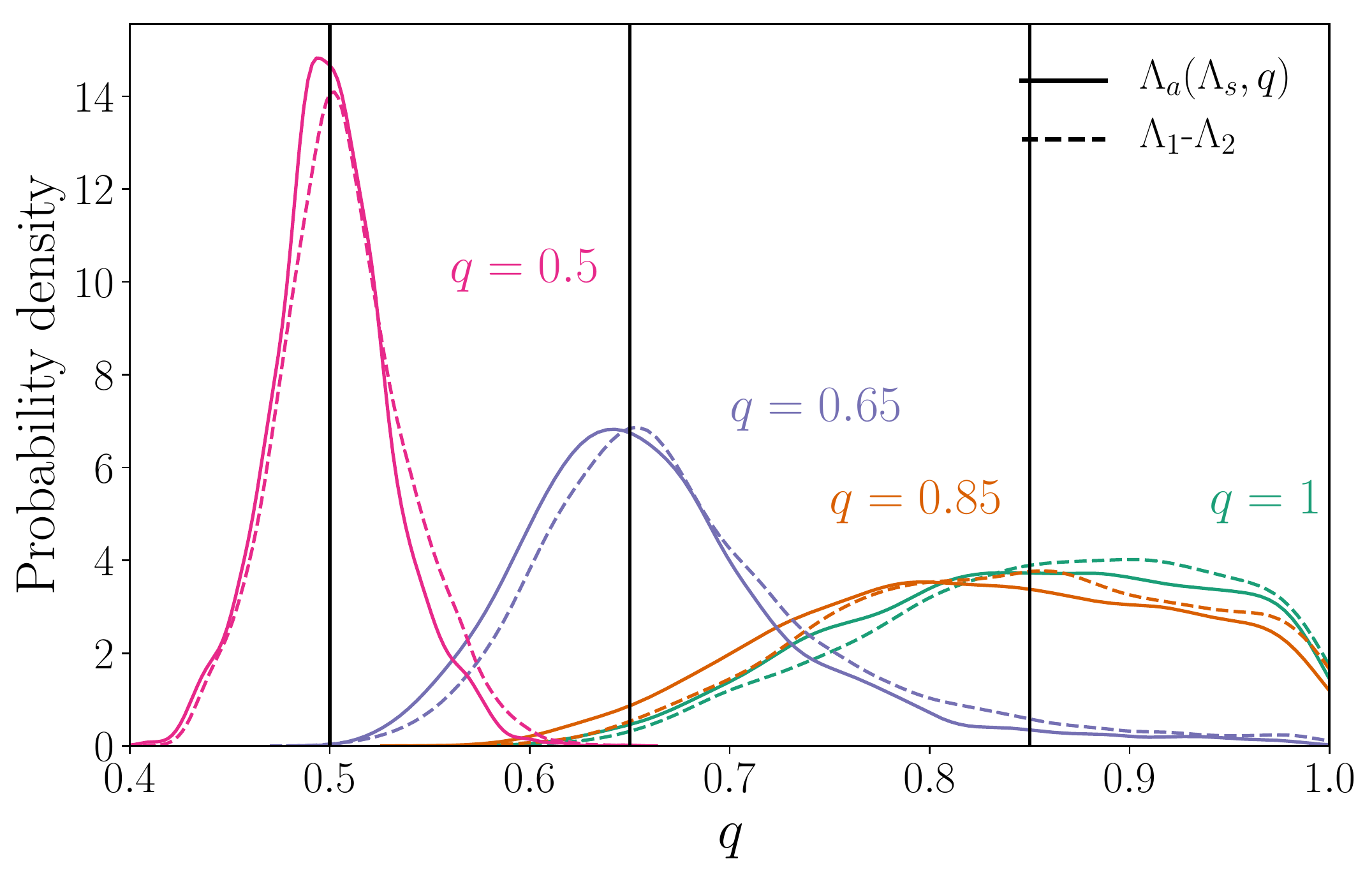}
\caption{ \label{fig:q_H4_SNR15} Marginalized posterior density for the mass ratio for H4, SNR 15 (top) and SNR 30 (bottom) when assuming that the individual tidal deformabilities are independent (dashed lines), and when we employ the EoS-independent relation (solid lines). The vertical black lines denote the injected values for the mass ratios (the $q=1$ line is not discernible).}
\end{figure}

Figure~\ref{fig:q_H4_SNR15} explores this by showing the posterior for the mass ratio for signals created with the H4 EoS and different SNRs and injected mass ratios. 
The top panel shows results for an SNR of 15, suggesting that the mass ratio is only mildly affected by the use of the EoS-independent relation (difference between solid and dashed lines of the same color). 
Once the SNR of the signal is increased to 30 (bottom panel) the mass ratio posteriors are very similar.
This indeed shows that the mass ratio is measured from a lower PN order in the GW phase than the tidal parameters.

\section{Conclusions}
\label{conclusions}

We have presented and studied a method to improve the measurement of tidal effects from GW signals. Our approach utilizes an EoS-independent relation between the tidal deformabilities of the two binary components given the ratio of their masses. The residual error in this relation is marginalized over so as to avoid any systematic biases in tidal inference. We have tested this method on simulated BNS signals for various EoSs and mass ratios, finding an improvement in the measurement of the individual tidal parameters by up to an order of magnitude, depending on the EoS and the mass ratio. Simultaneously, we find an increased evidence for the BNS model using the EoS-independent relation, suggesting that it facilitates distinguishing between BNS and NSBH systems.

Besides the systematic bias due to the use of the EoS-independent relation, parameter estimation of a real signal could potentially suffer from additional systematic effects due to any mismatch between our waveform models and the signals. In the case of simulated signals, such as the ones studied here, this is avoided by the fact that we use the same waveform model both for the creation and for the analysis of said signals. The waveform model we use, {\tt IMRPhenomD\_NRTidal}, has been shown to reliably reproduce results from numerical simulations of BNS mergers~\cite{Dietrich:2018uni}, however it still neglects a number of physical effects, such as spin-precession and the monopole-quadrupole interaction~\cite{Poisson:1997ha}. In the case of a real signal, a way to assess the effect of model inaccuracies is to perform analysis with a number of waveform models, as for example in Ref.~\cite{TheLIGOScientific:2017qsa}.

The monopole-quadrupole term, in particular, arises due to the interaction between the monopole of one NS and the spin-induced quadrupole moment of the other NS. Despite entering at a lower PN order than the tidal deformability (2PN) it is proportional to the spin squared, which is in general a small quantity given that NSs in binaries are expected to be spinning slowly. If this expectation is violated, then omitting this term could induce considerable biases~\cite{Harry:2018hke}, which is why in this study we employ the small-spin prior of Ref.~\cite{TheLIGOScientific:2017qsa}. We note that the effect of the monopole-quadrupole term could be seamlessly incorporated in our analysis through use of another EoS-independent relation, this time between the quadrupole moment and the tidal deformability of a NS~\cite{Yagi:2013bca,Yagi:2013awa}, as has already been done in a number of studies~\cite{Agathos:2015uaa,Chatziioannou:2015uea}.

Besides the EoS-independent relation between the individual tidal parameters and the mass ratio studied here, a number of other such universal relations have been proposed. One example is the relation between the tidal deformability and the quadrupole moment described above. A second example consists of a relation between the tidal deformability of a star and its compactness $C\equiv m/R$, where $R$ is the NS radius~\cite{Maselli:2013mva,Urbanec:2013fs}. Use of this relation would enable us to translate our results for the measurement of the tidal parameters to a direct measurement of the two-dimensional mass-radius posterior. In this study, we choose to incorporate only the relation between the individual tides, as we are interested in studying its implications without additional assumptions. Incorporation of additional universal relations (and marginalization of their intrinsic errors) is left for future study.

As a final remark, we note that the study presented here makes use of simulated GW signals, without any provision for realistic GW detector noise. In reality, the detector noise can violate the assumptions of stationarity and Gaussianity making inference less efficient. Indeed, this was the case for GW170817, whose signal coincided with an nonastrophysical transient, known as a glitch. The study presented here assumes that such obstacles can be efficiently dealt with in a realistic detection scenario similarly to GW170817~\cite{TheLIGOScientific:2017qsa,glitchmitigation}.

\acknowledgements
 
We thank Kent Yagi for fruitful discussions on the EoS-independent relations and for sharing his data with us. We thank Christopher Berry and Wynn Ho for helpful comments on the manuscript. The authors acknowledge the LIGO Data Grid clusters.

\bibliography{OurRefs}

\end{document}